\documentclass{aa}
\pdfoutput=1

\usepackage{txfonts}
\usepackage{xcolor}
\usepackage{graphicx}
\usepackage{array}
\usepackage{amsmath}
\usepackage{float}
\usepackage{hyperref}
\usepackage{ulem}
\usepackage{makecell}

\hypersetup{
    colorlinks,
    linkcolor={red!50!black},
    citecolor={blue!50!black},
    urlcolor={blue!80!black}
}
\bibpunct{(}{)}{;}{a}{}{,} % to follow the A&A style
\newcommand{\new}[1]{#1}
\newcommand{\nnew}[1]{#1}

\begin{document}

\title{\new{Analysis of the arm-like structure in the outer disk of PDS~70}}
\subtitle{\new{Spiral density wave or vortex?}}
    
    \author{S.~Juillard\inst{1}\fnmsep\thanks{F.R.S.-FNRS PhD Research Fellow} 
          \and
          V.~Christiaens\inst{1}\fnmsep\thanks{F.R.S.-FNRS Postdoctoral Fellow}
          \and
          O.~Absil\inst{1}\fnmsep\thanks{F.R.S.-FNRS Senior Research Associate}}
      
    \institute{STAR Institute, Universit\'e de Li\`ege, All\'ee du Six Ao\^ut 19c, 4000 Li\`ege, Belgium \\ \email{sjuillard@uliege.be}}
    
    \date{Received 02 July 2022, Accepted 2 November 2022}
    
    \abstract
    {Observing dynamical interactions between planets and disks is key to understanding their formation and evolution. Two protoplanets have recently been discovered within the  PDS~70 \new{protoplanetary} disk, along with an arm-like structure toward the northwest of the star.}
    {Our aim is to constrain the morphology and origin of this arm-like structure, and to assess whether it could trace a spiral density wave caused by the dynamical interaction between the planet PDS~70c and the disk.}
    {We analyzed polarized and angular differential imaging (PDI and ADI) data taken with VLT/SPHERE, spanning six years of observations. The PDI data sets were reduced using the \texttt{irdap} polarimetric data reduction pipeline, while the  ADI data sets were processed using \texttt{mustard}, a novel algorithm based on an inverse problem approach to \nnew{tackle the geometrical biases spoiling the images previously used for the analysis of this disk}.}
    {We confirm the presence of the \new{arm-like structure in all PDI and ADI data sets, and extract its} trace by identifying local radial maxima in azimuthal slices of the disk in each data set. \nnew{We do not observe a  southeast symmetric arm with respect to the disk minor axis, which seems to disfavor the previous hypothesis that the arm is the footprint of a double-ring structure}. If the \new{structure traces a} spiral \new{density wave following} the motion of PDS~70c, we \new{would} expect $11\fdg28^{+2\fdg20}_{-0\fdg86}$ rotation for the spiral in six years. However, we do not measure any significant movement of the \new{structure}.}
    {If the \new{arm-like structure} is a planet-driven spiral arm, the observed lack of rotation would suggest that the assumption of rigid-body rotation may be inappropriate for spirals induced by planets. We suggest that the arm-like structure may instead trace a vortex appearing as a one-armed spiral in scattered light due to projection effects. The vortex hypothesis accounts for both the lack of observed rotation and the presence of a nearby sub-millimeter continuum asymmetry detected with ALMA. Additional follow-up observations and dedicated hydrodynamical simulations could confirm this hypothesis.}


\maketitle

% \input{obs_summary_tab.tex} 
% _____________________________________
% Summary of the observations
\begin{table*}[ht] 
\begin{center}
\caption{Summary of the {\sc SPHERE/IRDIS} observations of PDS~70 used in this work, sorted by date.} 
\label{tab:obs}
\begin{tabular}{lcccccccccc}
\hline
\hline
Date & Strategy & Program & Filter & Coronagraph  & 
T$_{\rm int}^{\rm (a)}$ & number &seeing$^{\rm (c)}$ & $\Delta$PA$^{\rm (d)}$ \\
& & & & &
[sec] & of frames$^{\rm (b)}$ & [\arcsec] & [\degr] \\ 
\hline

2015-05-06 & ADI & \href{http://archive.eso.org/wdb/wdb/eso/sched_rep_arc/query?progid=095.C-0298(A)}{095.C-0298(A)}  & $\nnew{H23}$  & N\_ALC\_YJH\_S  &
64 & 16 & 1.25 & 36.0\\

2015-06-04 & ADI & \href{http://archive.eso.org/wdb/wdb/eso/sched_rep_arc/query?progid=095.C-0298(B)}{095.C-0298(B)}  & $\nnew{H23}$  & N\_ALC\_YJH\_S  &
64 & 16 & 1.33 & 40.0\\

2016-03-26 & PDI & \href{http://archive.eso.org/wdb/wdb/eso/sched_rep_arc/query?progid=096.C-0333(A)}{096.C-0333(A)}  & $J$  & N\_ALC\_YJ\_S  &
64 & 35 & 1.83 & -\\

2018-01-25 & ADI & \href{http://archive.eso.org/wdb/wdb/eso/sched_rep_arc/query?progid=1100.C-0481(D)}{1100.C-0481(D)}  & $\nnew{K12}$  & N\_ALC\_YJH\_S & 96 & 30 & 1.07 & 95.7\\

2019-03-11 & ADI & \href{http://archive.eso.org/wdb/wdb/eso/sched_rep_arc/query?progid=1100.C-0481(L)}{1100.C-0481(L)}  & $\nnew{K12}$  & N\_ALC\_Ks  & 96 & 16 & 0.7 & 49.0\\

2019-04-13 & ADI & \href{http://archive.eso.org/wdb/wdb/eso/sched_rep_arc/query?progid=1100.C-0481(M)}{1100.C-0481(M)}  & $\nnew{K12}$  & N\_ALC\_YJ\_S  &
96 & 15 & 0.53 & 71.3\\

2019-07-13 & PDI & \href{http://archive.eso.org/wdb/wdb/eso/sched_rep_arc/query?progid=1100.C-0481(T)}{1100.C-0481(T)}  & $Ks$  & n/a & 64 & [66,69]$^{\rm (e)}$ & [0.9,0.69] & -\\

2019-08-09 & PDI & \href{http://archive.eso.org/wdb/wdb/eso/sched_rep_arc/query?progid=0102.C-0916(B)}{0102.C-0916(B)}  & $H$  & N\_ALC\_YJH\_S  & 64 & 37 & 1.58 & -\\

2021-07-16 & PDI & \href{http://archive.eso.org/wdb/wdb/eso/sched_rep_arc/query?progid=60.A-9801(S)}{60.A-9801(S)}  & $H$  & \makecell{ \nnew{[N\_NS\_CLEAR},\\\nnew{N\_ALC\_YJH\_S]}}  &
16 & [17,49]$^{\rm (f)}$ & [0.65,0.83] & -\\

\hline
\end{tabular}
\end{center}
Notes: $^{\rm (a)}$ Exposure time of OBJECT frames. $^{\rm (b)}$ Number of frames in cubes. $^{\rm (c)}$ Average seeing from \href{https://www.eso.org/asm/ui/publicLog}{ESO Astronomical Site Monitoring (ASM)}. $^{\rm (d)}$ Maximum parallactic angle variation. \nnew{$^{\rm (e)}$ Program split into two parts to improve image quality, due to the variation in observing conditions}. $^{\rm (f)}$ Program split into two parts, \new{due to different coronagraph settings}. 
\end{table*}
% _____________________________________

\section{Introduction}
   
    PDS~70 is a young ($5.4\pm1$~Myr) T Tauri star hosting a protoplanetary disk and two nascent exoplanets \citep{PDS70info,muller,PDS_first_claim_planet_c}. The disk has an inclination of $49\fdg7\pm 0\fdg3$ and  a position angle of semimajor axis of $158\fdg6\pm 0\fdg7$, and features an inner cavity of radius $\sim$55~au \citep{Hashimoto, PDS70info}. The star was observed with high-contrast imaging instruments from 2015 to 2021 in both polarized differential imaging (PDI) and angular differential imaging (ADI) observing strategies \new{in the near-infrared (NIR)}. \new{Some of these observations displayed an asymmetrical feature to the northwest (NW) of the star, which was interpreted as the potential signature of a double ring \citep{PDS70info, muller, periodPlanetc} or of planet-disk interactions \citep{Holstein21}.} \new{In these studies image processing of ADI sequences was performed with classical techniques such as median- or PCA-based PSF subtraction, even though it is known to cause the deformation of extended sources \citep{Milli,Valentin}. More recently, three algorithms based on inverse-problem approaches have been developed to tackle the retrieval of extended signals in high-contrast imaging data sets: \texttt{mayonnaise} \citep[]{mayo}, \texttt{rexpaco} \citep[]{REXPACO}, and \texttt{mustard} (Juillard et al., in prep.). Using these new algorithms, the morphology of the asymmetrical feature appears as an arm-like structure, whose origin and connection with the global disk architecture remains to be determined}.
    
    In theory, this kind of arm-like structure can form when the orbital frequency of a planet embedded in a protoplanetary disk resonates with the epicyclic frequency of the disk gas proper motion \citep{Lindblad}. This phenomenon, called a  Lindblad resonance, creates constructive interference of density waves that leads to the formation of a primary (and potentially additional) spiral arm(s) \citep{Rafikov,Bae}. \new{Alternatively, single arm-like structures can be the result of projection effects. For instance, a slight differential tilt of the outer part of the disk can create illumination or shadowing effects such that the disk  can appear to display spiral arms in scattered light images \citep{shadowstudy}. Furthermore, anti-cyclonic vortices triggered by the Rossby wave instability at the edge of planet-opened gaps can also cause such arm-like patterns \citep{VotexMimicSpiral}.} A vortex can actually trap large quantities of dust, creating asymmetries in the density distribution within the protoplanetary disk \citep{vortex_formation, vortex_planet, vortex_no_planet}. In a flared and inclined disk\new{, an overdensity may mimic a one-armed spiral in scattered light images, due to perspective effects \citep{VotexMimicSpiral}, as potentially observed in the disks of HD~34282 \citep{deBoer21} or HD~143006 \citep{HD143006}}. Disk self-gravity \citep{selfgravity}, which is expected to originate mainly in very young and massive disks \citep{Long}, is another possible source of spiral structure, and does not match the case of PDS~70. \nnew{Similarly, the absence of observed shadows in the disk of PDS~70 rules out the hypothesis of a spiral triggered by the local cooling of the disk by a shadow \citep{Montesinos16}}. Finally, an arm-like structure could also be the result of the gravitational interaction between the disk and a massive nearby object during a flyby event \citep{Cuello2020,flyby}.
    
    Arm-like structures have been observed in a large number of protoplanetary disks  \citep[e.g., HD~100546, HD~141569, or HD~135344B;][]{spiral1, spiral2, spiral3}. To date,  none of these spirals arms has been associated with a confirmed exoplanet, hence their origin remains unclear as other physical mechanisms can lead to arm-shaped structures. \new{Given its proximity with planet c, the arm-like structure observed in the outer disk of PDS~70 might be directly connected to it}. This suggests that PDS~70 may offer the unique opportunity of studying \new{planet-disk interactions} in a case where the planet location is known.
   
    \new{In this work, we confirm the presence of an arm-like structure in the disk of PDS~70 using a new algorithm based on an inverse approach, referred to as \texttt{mustard}, and trace the evolution of its position and morphology over the six years of high-contrast imaging observations available on this source. In Sect.~\ref{sec:obs} we present the data and the post-processing techniques used in this study, and we compare our new method to previous studies. Section~\ref{sec:analysis} is dedicated to the analysis of the post-processed images, with specific analyses performed for three testable scenarios: a double-ring structure, a spiral density wave associated with PDS~70c, and a stellar flyby. %First, we characterize the arm-like structure by computing its trace in the nine coronagraphic data sets available to us. Second, we investigate the planet-driven spiral arm hypothesis by testing whether the arm-like structure follows the motion of planet c as a rigid body---the expected motion of a planet-driven spiral arm. Third, we analyse the position of a nearby point-like source, located at about 2\farcs54 from PDS~70, in an attempt to determine whether it could have created the arm-like structure through to a recent flyby \citep{flyby}. 
    In Sect.~\ref{sec:discussion} we discuss the likelihood of various physical mechanisms to be at the origin of the observed structure. We conclude our study in Sect.~\ref{sec:conclusion}.}

%--------------------------------------------------------------------

\section{Observations and data processing} \label{sec:obs}
\label{sec:reduction}

\subsection{Data sets and pre-processing}

    We used all publicly available coronagraphic \new{(and one non-coronagraphic)} ADI and PDI data sets of PDS~70 taken by  the InfraRed Dual Imager and Spectrograph (IRDIS) of the SPHERE high-contrast instrument installed on the Very Large Telescope \citep{sphere}. \new{A total of} five observing programs using the ADI strategy from 2015 to 2019 and four programs using the PDI strategy from 2016 to 2021, \new{were available (see Table~\ref{tab:obs})}.

    We used \texttt{irdap}\footnote{\url{https://github.com/robvanholstein/IRDAP}} \citep[IRDIS Data reduction for Accurate Polarimetry,][]{irdap} to reduce all the PDI data sets. The pipeline is highly automated and performs sky subtraction, flat-fielding, bad pixel removal, and centering \new{before calculating the  $Q_{\phi}$ and $U_{\phi}$ maps}. We used the default \texttt{irdap} parameters for the reduction except for the \texttt{annulus\_star} keyword, which was set to ``star aperture'' for the star polarization  to be determined within a small radius located at the position of the central star; this  allowed us to remove  the stellar and instrumental polarization (R. van Holstein, pers. comm.). \new{The PDI post-processed images are displayed in Appendix~\ref{sec:gallery}.}
    
    The ADI data sets were pre-processed using \texttt{vcal\_sphere}\footnote{\url{https://github.com/VChristiaens/vcal_sphere}} \citep{Valentin2}. This pipeline makes use of the Vortex Image Processing   (\texttt{VIP}\footnote{\url{https://github.com/vortex-exoplanet/VIP}}) \citep[][]{vipref} package, which  provides processing tools for high-contrast imaging, and of the ESO Recipe Execution Tool \texttt{EsoRex}\footnote{\url{https://www.eso.org/sci/software/cpl/esorex.html}}. It performs flat-fielding, PCA-based sky subtraction, bad pixel correction, bad frame removal, and centering of the images. The centering was done in two steps: a coarse centering based on satellite spots and a fine centering based on the expected circular motion of a bright background star located at an angular separation of 2\farcs54, with a position angle of 14\fdg7.

    \subsection{\new{ADI cube processing with \texttt{mustard}}}
    
    \new{Median- and PCA-based post-processing methods were used in previous studies presenting ADI data sets on PDS~70 \citep{PDS70info,muller,PDS_first_claim_planet_c}. The main limitation of these methods is the poor handling of the circularly symmetric component of the extended signal, which is known to induce geometrical biases in the post-processed images \citep{Milli}. This limitation is inherent to the ADI strategy: circularly symmetric components are invariant to field rotation, and thus appear static in the image data cube, similarly to the stellar halo that needs to be subtracted from the images.} \new{To mitigate this effect we designed a new algorithm called \texttt{mustard} (Juillard et al., in prep.), which follows an inverse-problem approach similar to the \texttt{mayonnaise} \citep[]{mayo} and \texttt{rexpaco} \citep[]{REXPACO} algorithms. We summarize the main steps of the algorithm below, and leave the  details of the  extensive tests applied to a variety of synthetic disks to a forthcoming paper.}
    
    \new{The strategy can be described as a search for the 2D maps $[d, s] \in \mathcal{R}^{+\; [m,\,m]}$, of size $m\times m$ pixels, representing respectively the circumstellar signal $d$ (positive rotating contribution) and the stellar PSF $s$ (positive static contribution) that minimize the distance of the ADI cube to a composite model of the ADI cube:}
%    \begin{figure}%[H]
%        \centering
%        \includegraphics[width=\linewidth]{images/MUSTARDb.png}
%    \end{figure}
    \begin{equation}
        J=\underset{s , d \in  \mathbf{R}^{+ [ m\times m]}}{\overset{}{\arg\min}}   \sum _{\mathrm{pixels}}^{m\times m}\left[ \mathrm{M}_{c}  \times \left(  \sum _{k=0}^{n}[  Y\boldsymbol{_{k}} -(s+R_{\theta _{k}}( d))  ]^{2}  \right)  + P_{1} +P_{2}\right] \, . \label{eq:argmin}
    \end{equation}
    \new{Here $\mathrm{M}_c$ is a numerical mask representing the coronagraph; $Y_k$ is the k-th frame of the ADI cube, associated with a parallactic angle $\theta_k$,  where the rotation operator for an angle $\theta_k$ is written $R_{\theta_k}$. The priors $P_1$ and $P_2$   are respectively defined as 
    \begin{equation}
        P_{1}  = \mu _{1}(\mathrm{M}_{R} \times d) \, , \\
        P_{2}  =\mu _{2}\left(\frac{\delta d}{\delta \mathrm{x} \delta \mathrm{y}} +\frac{\delta s}{\delta \mathrm{x} \delta \mathrm{y}}\right)   \, , 
    \end{equation}
    where $\mu_1$ and $\mu_2$ are scalar weights, and $\mathrm{M}_R$ is a 2D mask. The algorithm is initialized with $d$ equal to the median of the de-rotated cube, so that all circularly symmetric flux is initially assigned to the circumstellar signal map. Then our current solution for handling circularly invariant signals consists in defining a mask $\mathrm{M}_R$, and   using it in a regularization term $P_1$ to force the circularly symmetric flux close to the star to be  associated with the stellar halo, and the circularly symmetric flux outside the mask to instead be associated with the disk. In the case of PDS~70, the presence of a cavity helps in adjusting our prior by defining the mask as a circle of $0\farcs18$ radius  (i.e., slightly smaller than the cavity radius projected along the disk minor axis) in order to capture most of the stellar halo but not the disk signal. The other regularization term ($P_2$) aims to leverage the correlation between neighboring pixels via the computation of the spatial gradient of $[d, s]$. %\sout{Additionally, the principal criterion is enhanced with strategies to compensate for other potential sources of errors including uneven repatriation of the frames rotation angles ($cw_k$), potential variations of intensity ($A_k$), and technical bottlenecks such as the known interpolation bias of the rotation operator of 2D images in the discrete domain.} 
    The weights $\mu_1$ and $\mu_2$ of the regularization terms $P_1$ and $P_2$ are computed automatically so that they account for about $5\%$ of the overall expression to be minimized in Eq.~\ref{eq:argmin}. Based on the tests performed so far (to be described in a forthcoming paper), \texttt{mustard} preserves the flux of the disk significantly better than classical algorithms such as median- or PCA-based PSF subtraction, and restores the image of disks obtained in ADI with more fidelity (see Sect.~\ref{sec:PCAbias}).} 
    
    \begin{figure}[!t]
    \centering
    \includegraphics[width=\linewidth]{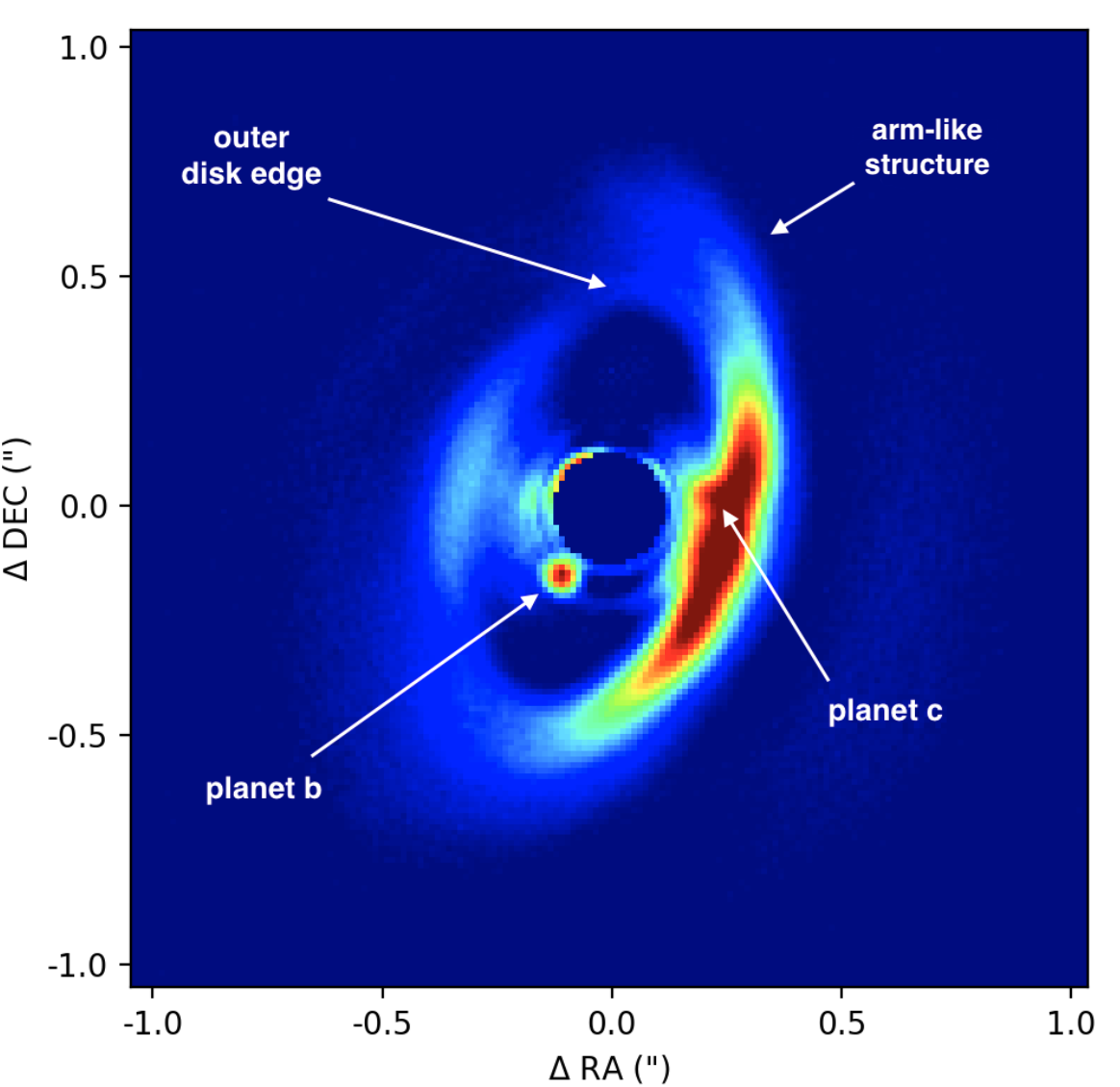}
    \caption{Final post-processed image for the ADI observation from March 2019 in the $K2$ filter (program 1100.C-0481(L)), based on the \texttt{mustard} algorithm. The intensity scale is linear. A central mask 16 pixels in diameter (0\farcs196, slightly larger than the Lyot mask of the SPHERE coronagraph) was added to hide the bright stellar residuals around the coronagraphic mask in the \texttt{mustard} post-processing.}
    \label{fig:exempleADI}
    \end{figure}
    
    An example of a PDS~70 ADI data set processed with \texttt{mustard} is shown in Fig.~\ref{fig:exempleADI}, \new{where the arm-like structure clearly extends outside of the disk outer edge to the NW. The rest of the ADI post-processed images are displayed in Appendix~\ref{sec:gallery}.} \new{Three} of our \new{five} ADI sequences were affected by a strong wind-driven halo \citep{Faustine2}, which \texttt{mustard} is unable to remove because it shares the same signature as the signal of interest (both the disk and the wind-driven halo rotate with the field of view). However, the wind-driven halo did not cover the outer parts of the disk, so  these three data sets could still be used to study the \new{arm-like structure}. \new{This is illustrated in Fig.~\ref{fig:ADIreduced}, where the  areas corrupted by the wind-driven halo are highlighted by hashed lines.}
    
    \begin{figure*}[!t]
    \centering
    \includegraphics[width=\linewidth]{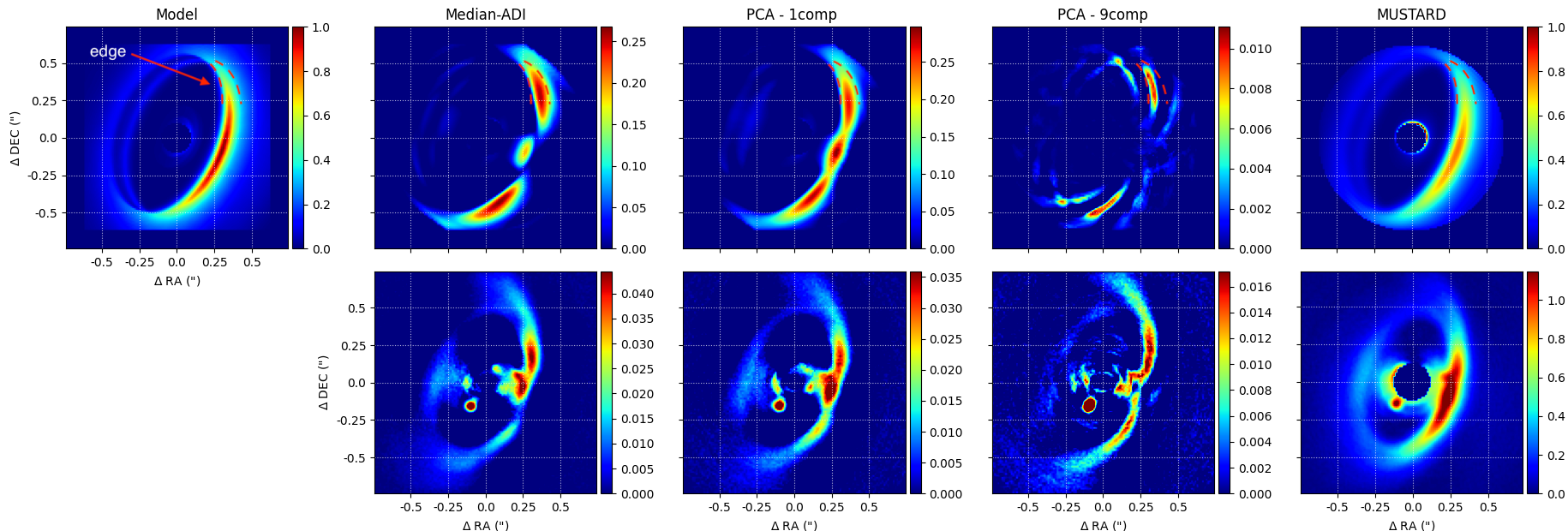}
    \caption{Comparison of different post-processing methods applied to fiducial and observed ADI sequences of PDS 70. \new{\textit{Top}: Final image obtained after post-processing an ADI sequence containing only a disk model (left) with different methods: median-ADI, PCA-ADI considering one and nine principal components, and \texttt{mustard} (from left to right). \textit{Bottom}: Application of the same post-processing methods to the ADI data set from March 2019 obtained in the $K$1 filter (program 1100.C-0481(L)). The axis grids of both model reductions and observation are scaled identically. The color axis represents the raw flux restored from reductions.}}
    \label{fig:PCA_vs_MUSTARD}
    \end{figure*}
    
    \subsection{\new{Validation of \texttt{mustard} on a disk model}}
    \label{sec:PCAbias}
    
    \new{In order to test the capability of \texttt{mustard} to properly extract the disk structure from an ADI sequence for a representative disk morphology, we performed ADI processing on a cube containing only a rotating radiative transfer model of the PDS~70 protoplanetary disk.} \new{The radiative transfer model is created with MCFOST \citep{MCFOST,Pinte2009}, using $1.3 \times 10^6$ Monte Carlo photon packets. To reproduce the pre-transitional disk morphology, we considered three zones in our radiative transfer model: an inner disk, a depleted annular gap, and an outer disk, with total dust masses of $10^{-9}$, $10^{-13}$, and $10^{-6} M_{\odot}$, and radially extending from 0.1 to 20~au, from 20 to 50~au, and from 50 au to 90~au, respectively. \nnew{The choice of disk mass and radial range for each zone is motivated by the closest visual match between predicted and observed PDS~70 images, among the  different sets of parameter values close to the ones suggested in \citet{Dong12} to reproduce the measured SED}. We assumed the same inclination, position angle (PA), and distance as in \citet{PDS70info}. Mie scattering is considered, and a grain size ($a$) population extending from 0.001 to $1 \mu$m is considered such that $n(a) \propto a^{-3.5}$. The model is computed for the $K1$ band ($\lambda = 2.11$~$\mu$m), and convolved with a Gaussian kernel whose full width at half maximum (FWHM) is set to 4 pixels, with a pixel scale of 12.25~mas (i.e., comparable to the FWHM and pixel scale of the SPHERE/IRDIS observations considered here). After convolution, the predicted total intensity image is duplicated and rotated a number of times so as to reproduce an ADI cube spanning the same parallactic angle variations as in the observations presented in \citet{muller}}.
    
    \new{Our synthetic ADI cube is then processed with various post-processing methods, and compared with the post-processed images obtained from the actual observations (see Fig.~\ref{fig:PCA_vs_MUSTARD}). As expected, the median- and PCA-based PSF subtraction algorithms remove the circularly symmetric part of the disk, leading to various artificial structures in the post-processed images. In particular, the deformed image of the disk edge partly overlaps with the arm-like structure that we aim to study, which is one of the reasons why to date  this structure has not been given more attention in the literature. Conversely, the \texttt{mustard} algorithm provides an almost unbiased image of the injected disk, with a clean disk outer edge.} 
    %is partly blended with  with the \sj{convoluted disk edge} which created misleading patterns.} \sj{Is it better ? I want to say that it would be less blend that much is the edge was more sharpe}
    \new{Consequently, what appears as a double asymmetrical signal with median- or PCA-based ADI processing, showing on both the NW and the southeast (SE) sides of the disk, is found to be more consistent with a combination of the illuminated edge of the cavity and a single-arm shaped feature on the NW side of the disk in the \texttt{mustard} image. This conclusion is corroborated by \texttt{mayonnaise} \citep{mayo} and \texttt{rexpaco} \citep{REXPACO} processing of some of the PDS~70 data sets presented here.}
    
%    \vc{I'd remove the following (somewhat off-topic):}
%    \sout{Note that depending on the geometrical configuration of the disk surface and the grain sizes the light scattered from the disk will not be evenly polarized, the observation from PDI can also causes a bias toward the disk morphology}\citep{PDIbiais}.
    
%---------------------------------------------------------------

\section{Analysis of the processed images} \label{sec:analysis}

\new{We \new{re}-detect an asymmetrical, arm-like structure in the NW outer part of the disk in all our data sets (see Fig.~\ref{fig:exempleADI} and Appendix~\ref{sec:gallery}). In the post-processed ADI images provided by \texttt{mustard}, the feature is clearly distinguishable  extending away from the bright illuminated edge of the cavity. Similarly, a clear arm-shaped signal in the PDI images can also be observed toward the NW without a SE equivalent (see Appendix~\ref{fig:PDIreduced}). In this section we further  analyze the post-processed images in an attempt to constrain the origin of this structure. More specifically, we explore three possible origins for the structure that are associated with testable hypotheses in direct imaging data: a double-ring scenario, a spiral density wave associated with PDS~70c, and a gravitational perturbation (flyby) by a nearby off-axis point-like source.}

\subsection{\new{Characterization of the disk asymmetry}}
\label{sec:characterisation}

\begin{figure*}[!t]
    \centering
    \includegraphics[width=\linewidth]{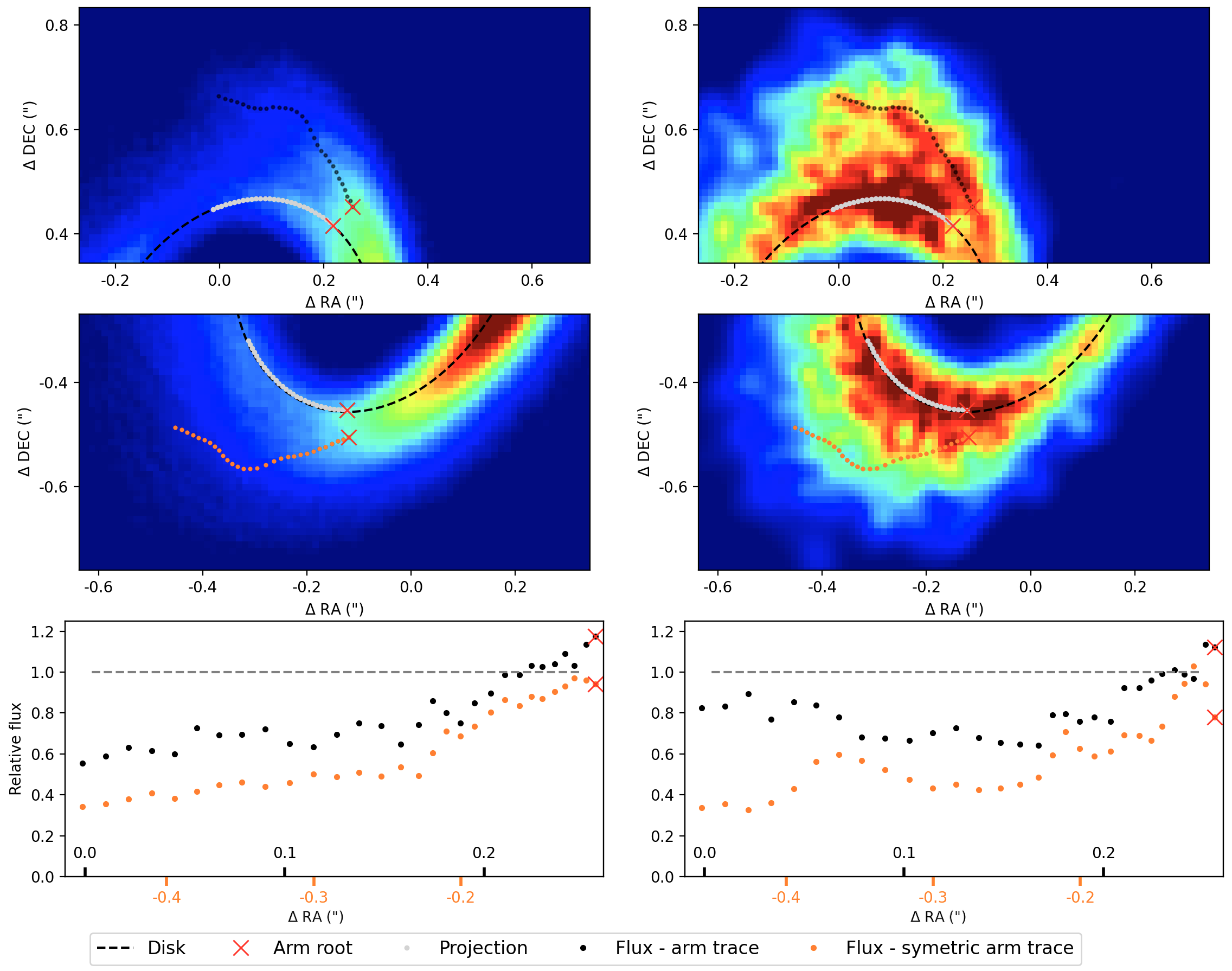}
    \caption{Comparison of the north (\textit{top}) and south (\textit{middle}) parts of the disk for 2019-03 ADI (\textit{left}) and 2019-06 PDI IRDIS (\textit{right}) observations obtained in $K$ band. The NW spiral  trace corresponds to the one measured on the ADI data set, following the method presented in Sect.~\ref{sec:characterisation}. The SE trace corresponds to the symmetric arm trace relative to the semimajor axis of the inclined disk plane, using the \citep{ALMAKeppler} disk parameters. The relative flux displayed in the bottom panels was computed on a four-pixel aperture. Each point is weighted by the flux measured in the disk at the same angle coordinate. The disk ellipse (dashed line) was computed to fit the local maxima found on the ADI image. The red and blue x-axis labels in the relative flux panel correspond respectively to the position of the NW arm trace relative to the star, and to the position of its SE symmetric.}
    \label{fig:comp_N_S}
\end{figure*}

\new{The morphology of the arm-like structure detected at the NW outer edge of the disk can be further characterized by computing its trace from each data set based on local radial maxima. In order to do so, we fitted a Gaussian to the radial intensity profile in azimuthal steps of 1\degr, and considered the center of the Gaussian. To refine our estimation of the trace, we applied a Laplacian filter (\new{i.e., a spatial second-order gradient) that highlights edges in an image.} We conservatively consider the uncertainty on each point of the trace to be the maximum between the uncertainty on the centroid of the Gaussian and 1/4 of the FWHM of the unsaturated PSF for each data set. The measurements were made after Laplacian filtering in order to reduce noise and to enhance the separation between the edge of the cavity and the spiral. Appendix~\ref{sec:laplace} illustrates with two examples how Laplacian filtering affects the trace measurement. The average spiral pitch angle over the spiral trace is measured to be $24\fdg4 \pm 2\fdg4$ after deprojection considering the disk orientation \citep{PDS70info}.}

\new{In order to test whether the extracted disk images could be compatible with a double ring structure, which should create a similar signature on the NW and SE sides of the disk, we characterize the asymmetry between the two sides of the disk using the two best-quality ADI and PDI images (see Fig.~\ref{fig:comp_N_S}). More specifically, we compare the flux along the trace of the NW arm-shaped structure to the flux in the SE mirrored image of the trace with respect to the minor-axis of the disk. 
The flux of the trace is weighted by the flux measured at the illuminated edge of the cavity at the same position angles, with these locations found by fitting an ellipse to local radial maxima.
%The flux of the trace is weighted by the intensity measured in the main disk at the same position angle, where the main disk is defined by fitting an ellipse to the local maxima detected in ADI images. 
In addition to the clear morphological difference between NW and SE noted in the post-processed image, we see in Fig.~\ref{fig:comp_N_S} that the flux of the trace relative to the disk is significantly larger on the NW side than on the SE: in the NW, the trace flux is equivalent to 81\% of the disk intensity at the same position angles on average, and is always above 55\% in the  ADI 2019-03 image and 69\% in the  PDI 2019-03 image, while in the SE, the symmetric trace flux represent only 61\% of the disk in average, going down to 33\% in both the  ADI and PDI images.}

\subsection{Measurement of arm motion over six years}

\begin{figure}[!t]
    \centering
    \includegraphics[width=\linewidth]{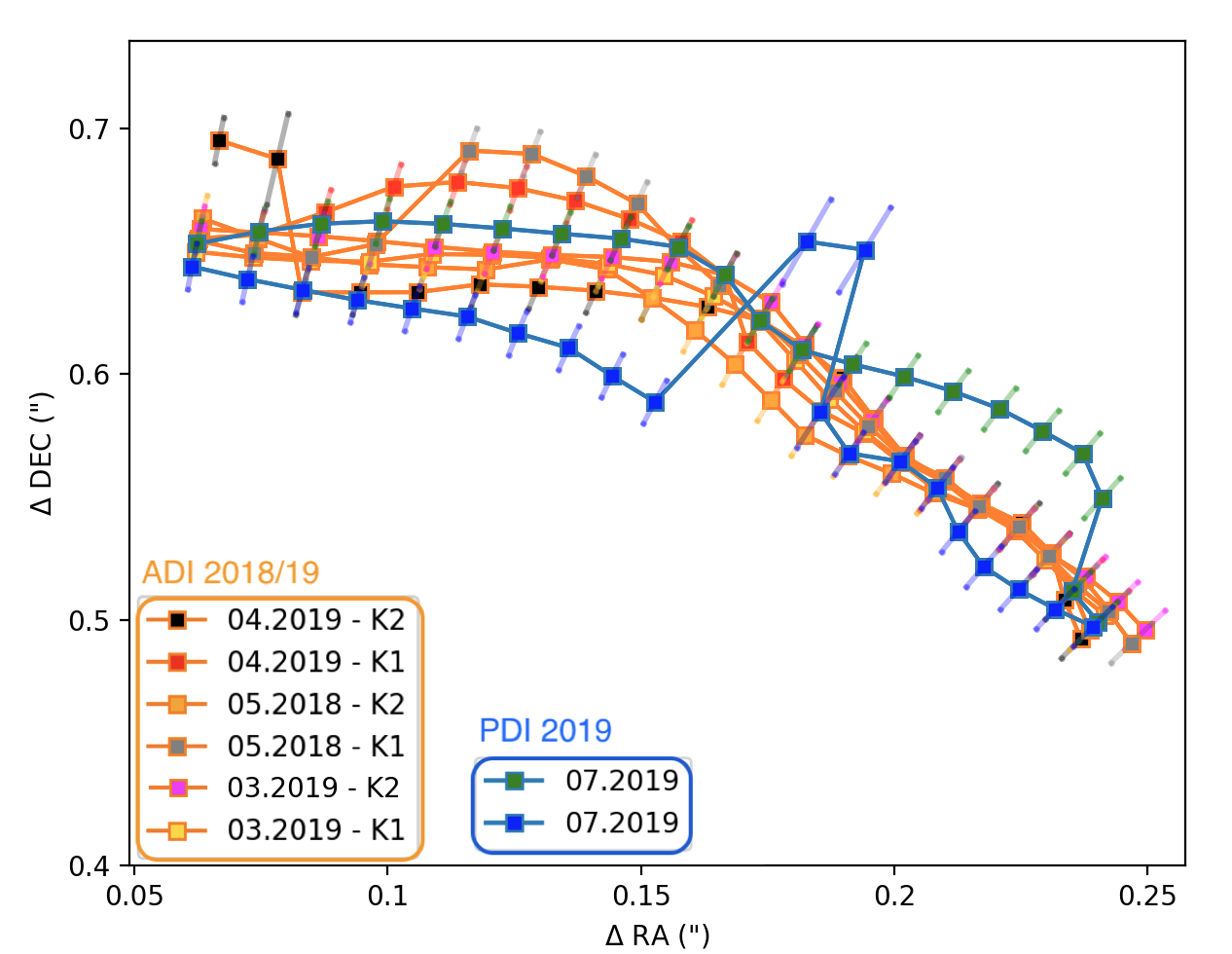}
    \caption{Spiral traces measured with local radial maxima at $K$ band in 2018 and 2019, for ADI-based (yellow-red tones) and PDI-based (blue-green tones) observing strategies.}
    \label{fig:spifit}
\end{figure}

\new{An intriguing scenario to explain the presence of the arm-like structure would be a planet-driven spiral density wave. In that scenario the observed structure would be expected to move in the disk at the same orbital velocity as the planet.} Protoplanet c has an orbital period of $191.5^{+15.8}_{-31.5}$ years \citep{periodPlanetc}. \new{Assuming the eccentricity to be negligible \citep[$e \simeq 0.05$,][]{Wang21},} we expect planet c to move by $11\fdg3^{+2\fdg2}_{-0\fdg9}$ in terms of position angle over six years. We computed the spiral shift prediction assuming a rigid body motion in the orbital plane of the planet, which means that the angular velocity is constant for every point of the arm-like structure, and assuming a circular orbit (the influence of the orbital eccentricity is negligible within error bars). As a consistency check, and in order to take into account the disk height, we considered a second method in which we deprojected the images with \texttt{diskmap}\footnote{\url{https://github.com/tomasstolker/diskmap}} \citep{diskmap} before identifying the local maxima, using the following parameters: $d=113.43$~pc, $i=49\fdg7$, PA=$158\fdg6$, $h(r) = 0.1(r/1{\rm au})^{1.25}$ \citep{PDS70info}. The trace identified in this way was then rotated and re-projected. We observed no significant difference between the two methods. The effect of disk height is smaller than our measurement uncertainties, meaning that in this particular case the disk height can be neglected.

\begin{figure}[!t]
    \includegraphics[width=\linewidth]{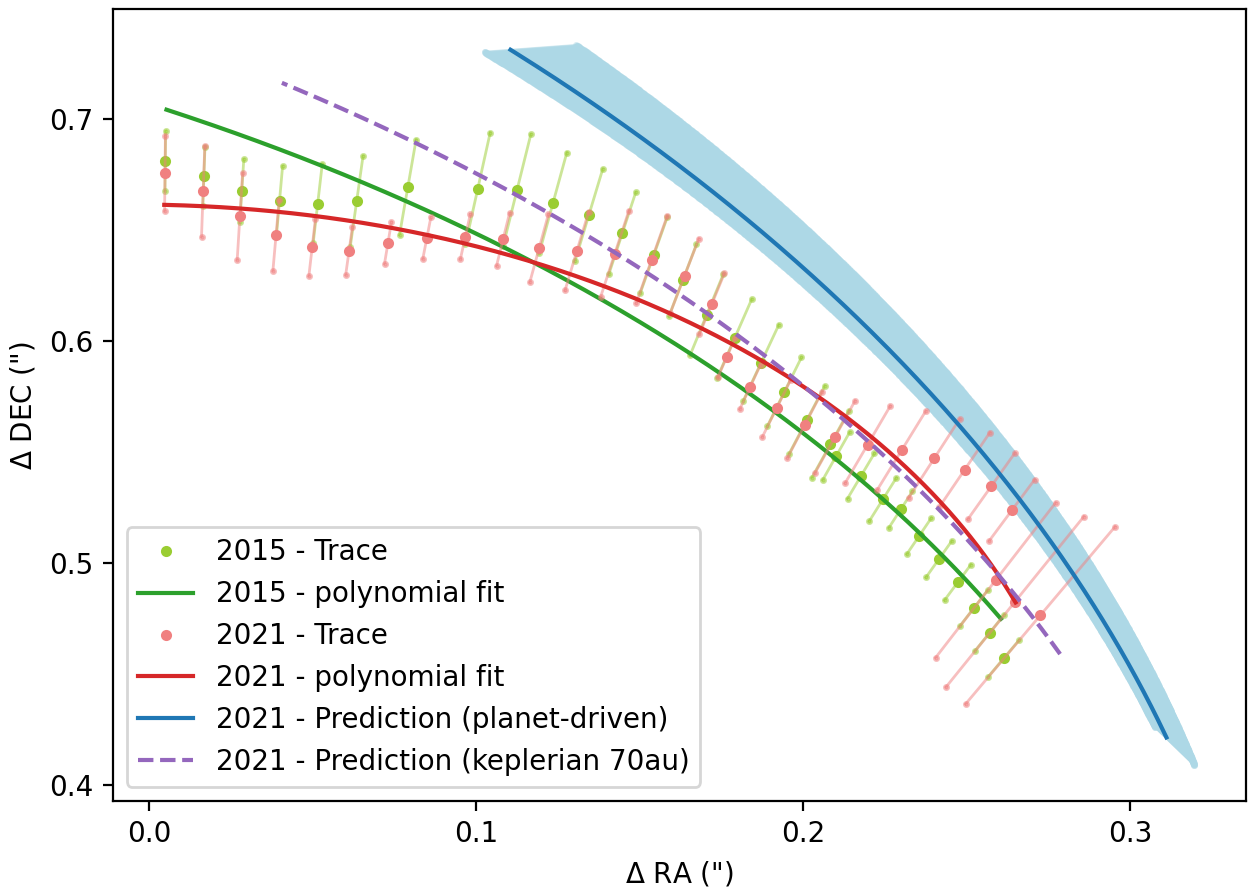}
    \caption{Comparison of the traces of the NW arm-like structure measured in the 2015 \new{(green)} and 2021 \new{(red)} data sets, with the \new{predicted location of the trace in 2021}, assuming respectively a rigid-body rotation associated with planet c (solid blue) and  a local Keplerian velocity at 70~au separation (dashed blue). The points with error bars correspond to the mean radial position measured at each azimuth on all data sets acquired in 2015 (two ADI data sets) and 2021 (two PDI data sets), while the solid lines correspond to a fit of the data points with a third-order polynomial. The error bars take into account both the standard deviation of the measurements and the radial uncertainty of the Gaussian fits used for the local maxima detection. The blue shaded region around the predicted 2021 trace corresponds to the uncertainty on the planet orbital parameters of \citet{periodPlanetc}.} %\todo{- Figure 3. I am not sure I understand the plots. Is the cyan lines the predicted location for the spiral in 2021 based on the 2015 trace in case of planet-driven (left) and keplerian rotation (right)? While at the third read I am sure about the answer, authors may want to be more clear about this in both the text and the caption to avoid other readers to lose their way like me. Also, note that the references to left and right plots may be swapped in the caption.}}
    \label{fig:obs}
\end{figure}

\new{Before exploring the possible motion of the arm-like structure, we compare the traces obtained with different observing strategies (ADI vs.\ PDI) from the same epoch.} This is an important check to make, as our oldest data sets (2015) and our most recent one (2021), both in $H$ band, were obtained with different observing strategies.
To perform this comparison, we sorted the science observations by band as the spiral morphology may be wavelength dependent \citep[e.g.,][]{MWC_wavelenght_dependecies}. In Fig.~\ref{fig:spifit} we plot the extracted trace for six ADI and two PDI reductions acquired close in time (2018/2019) in the $K$ band, with the aim of assessing whether the position of the trace is affected by a systematic bias due to the use of different observing strategies. The wavy aspect of the \new{local-maxima traces}, especially in the PDI images, is due to low-frequency noise (spatially correlated noise) that could not be corrected with the Laplacian filter. \new{This source of uncertainty is not accounted for in the Gaussian fit, which may therefore produce underestimated error bars. We also note that the 2019-07 PDI data set corresponds to the only non-coronagraphic data set among all considered observations, hence to a shorter total integration on target and a lower signal-to-noise ratio for the arm-like signal. This can also account for the   wavier aspect of the trace for that data set. Nonetheless no systematic bias is observed between PDI and ADI. As the minor differences observed between the traces inferred in the PDI and ADI observations are significantly smaller than the expected shift (see Fig.~\ref{fig:obs}), we conclude that exploring the possible movement of the arm-like structure with various observing strategies is appropriate}. %\sj{Answere to Olivier : So the problem is that it was a comment from the refere. J'ai essayé de réécrire la fin, car Olivier ne l'a pas bien comprise. } \todo{Referee :  I am not sure on how to interpret Figure 2. In the text, authors conclude that there is no shift between different datasets acquired in 2018/2019. However, the traces plotted in Figure 2 are not entirely consistent with each other, or at least some of them are not. Can authors clarify this point in case I am misunderstanding?} %\sj{In ADI images, as less light is reflected by the disk the further away from the outer disk}, the uncertainties for $\Delta RA < 0\farcs19\deg$ \sj{are probably underestimated.}

Figure~\ref{fig:obs} shows that a significant shift of $0\farcs048^{+0\farcs009}_{-0\farcs003}$ ($\sim$3.95~pixels) is predicted on average along the trace between 2015 and 2021, if the trace is co-moving with PDS~70c. We considered the arm-like structure to move as a rigid body, as expected from hydro-dynamical simulations \citep{Bae19, Toci20}. Our observations do not match this prediction. Instead, we observe no visible motion, with an average shift between the 2015 and 2021 traces smaller than $0\farcs012$ ($\sim 1$ pixel). Figure~\ref{fig:obs} also shows the prediction for Keplerian motion at the orbital distance of the arm-like structure itself, assumed to be $\sim$70~au. In this case the uncertainties are too large compared to the expected shift to conclude whether the trace is static or follows  Keplerian motion. 

%\new{Morover} it is important to highlight that the prediction does not take into account \sout{projection} effects \sj{due to the way the light is reflected by the disk} or other phenomena. Hence, it should be interpreted as an upper limit for the amplitude of the motion.

\subsection{\new{Astrometric and photometric analysis of a nearby star}} \label{sec:flyby}

\begin{figure}[!t]
    \centering
    \includegraphics[width=\linewidth]{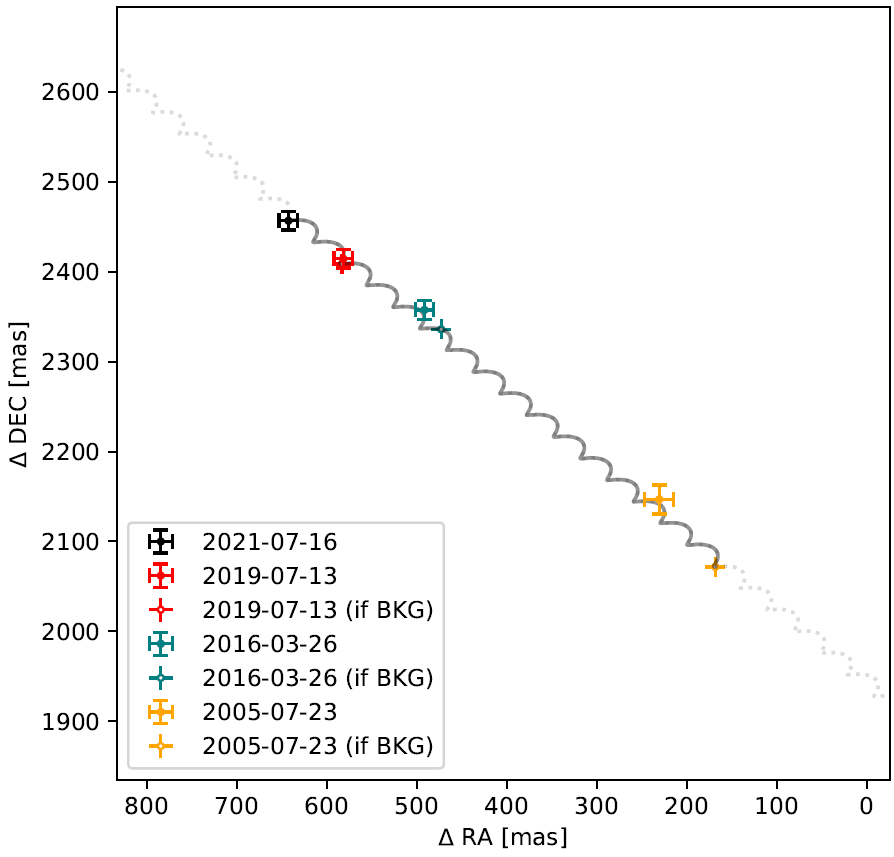}
    \caption{Relative position of the background star candidate (crosses with caps) near PDS~70 from the PDI data sets (see Table~\ref{tab:obs}) and the NACO data set presented in \citet{NACO_PDS}. Crosses with no caps show the expected positions (backward predictions based on the measurement from 2021) as estimated from the proper motion of PDS~70 with Gaia.}
    \label{fig:bkgstar}
\end{figure}

\new{The last testable hypothesis based on the available data relates to a possible gravitational interaction with the nearby point source already detected in previous studies \citep[e.g.,][]{NACO_PDS}, whose physical association with PDS~70  to
date has not  been formally concluded upon in the literature. In order to do so,} we performed an astrometric analysis of this point source, located at an angular separation $r=2\farcs54$ and position angle ${\rm PA}=14\fdg7$ from PDS~70, using multiple epochs spanning 16 years of observations, and considering the proper motion of PDS~70 measured by \citet{Gaia, Gaia2}. We considered the positions measured in our PDI data sets and re-investigated the 2005 observation of PDS~70 presented in \citet{NACO_PDS}, taken with the VLT/NACO instrument \citep{Lenzen, Rousset}. The comparison of the measured positions to the expected positions for a fixed background star (i.e., only considering the proper motion of PDS~70) shows a good, albeit imperfect, agreement (Fig.~\ref{fig:bkgstar}). \new{This result suggests that the off-axis star is not physically associated with PDS~70, although it may have a significant apparent motion of about 1/6th of the PDS~70 proper motion.}

\nnew{Furthermore, the Gaia DR3 estimated parallax of the point source ($0.042 \pm 0.038$ mas) suggests that it is located beyond $\sim$6.4~kpc at a 3$\sigma$ confidence level.  % no need to give too many significant digits given the amplitude of the uncertainties 
We also estimated the distance based on the measured $H-K$ color of the object in relevant data sets considered in this work, and assuming it corresponds to a main sequence star. The estimated color suggests a spectral type of A0 or earlier, which places the star at a large distance compatible with the lower limit inferred by Gaia. This confirms the background star hypothesis.}
%which locate the star at least 56 times farther from Earth than PDS~70. 

% Furthermore, we also estimated the $H-K$ color of the off-axis star to be $\sim 0.23$ based on the photometry measurements in our ADI data sets, which suggests it is an M4-type star \citep[Table.~7.6]{color}. Combining our magnitude measurements with the expected absolute magnitude of M4-type stars, we estimate the star to be about 10 times farther from Earth than PDS~70 ($\sim 1017$~pc; $\sim23722$~pc from Gaia GR3 database), which confirms the background star hypothesis.

%We conclude that it is a background star, which cannot dynamically interfere with the disk.

% ------------------------------------------------------------------------

\begin{figure*}[!t]
\centering
\includegraphics[width=\linewidth]{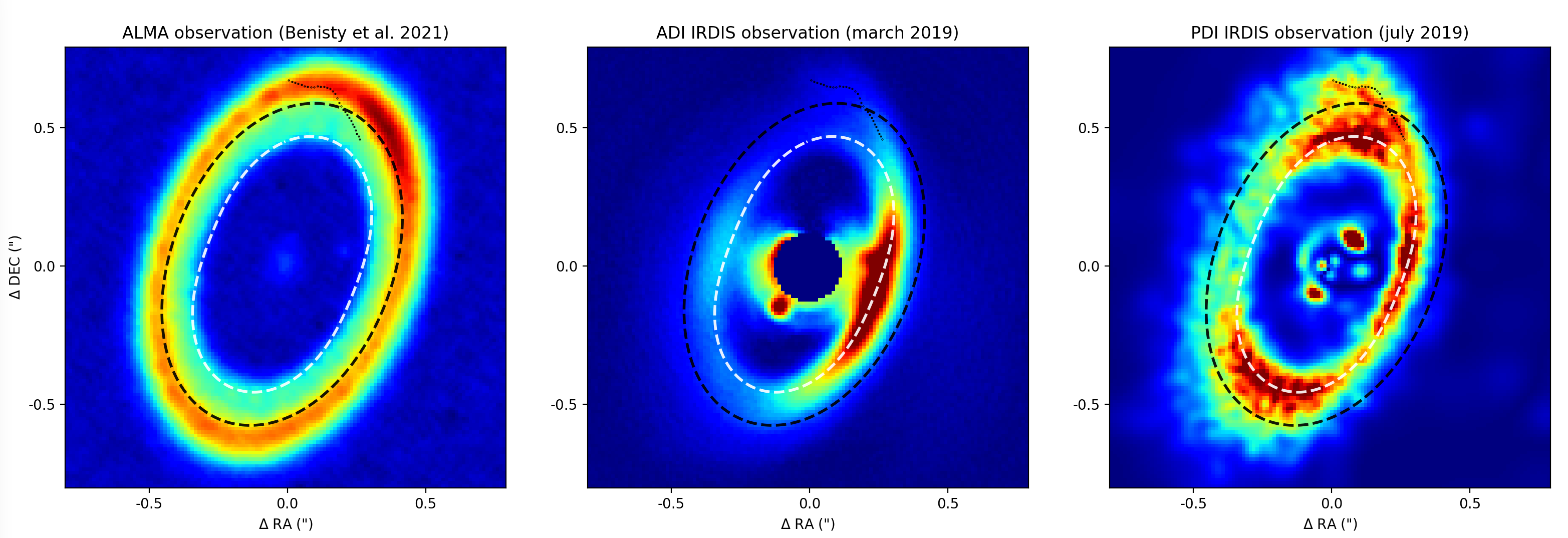}
\caption{ALMA sub-millimeter continuum observation presented in \cite{Myriam} compared to the 2019-03 ADI and 2019-06 PDI IRDIS observations in $K$-band. The spiral trace corresponds to the one measured on the ADI data set (method presented in Sect.~3.1). The dashed ellipses are computed as a circle on the inclined plane of   PDS~70. The radius and center have been adjusted to match the edge of the inner ring or pedestal (white), and of the outer ring (black), respectively, on the ALMA image.}
\label{fig:ALMAobs}
\end{figure*}

\section{Discussion} % since we are aiming for a letter, this is merged with the conclusions
\label{sec:discussion}

In the previous section we performed specific analyses of our post-processed images to assess their compatibility with various scenarios that could explain the presence of an arm-like structure to the NW of the star.
%\new{We analyzed morphological biases caused by PCA and median-ADI reductions using a toy model of the PDS~70 disk (Fig.~\ref{fig:PCA_vs_MUSTARD}). We noted that these post-processing methods yielded an arm-like shaped feature partly blended with the widened disk edge. The novel algorithm \texttt{mustard} enables more accurate recovery of the feature, showing a clear arm-shaped structure extending away from the illuminated edge of the cavity.}
\new{In} the first high-quality ADI images of PDS~70 presented in \citet{PDS70info} and \citet{muller}, the structure of the outer disk image was interpreted as the signature of a double-ring structure.
Moreover, the sub-millimeter continuum observed with ALMA (Fig.~\ref{fig:ALMAobs}) was successfully modeled with a pedestal-shaped ring that can be interpreted as an unresolved double ring \citep{Myriam}. 
We therefore started our exploration of various scenarios by comparing the brightness of the arm-like structure to the main disk, and performing a similar analysis on the mirrored SE side of the disk. This visual and quantitative analysis is enabled by our new post-processing method (\texttt{mustard}), which preserves the disk structure extracted from ADI data sets, and does not create strong artifacts, such  as those created using the standard median- or PCA-based algorithms. Our analysis shows that an arm-like structure is not detectable on the SE side of the disk or at least that, if present, its brightness is significantly reduced compared to the NW side. This does not preclude the possibility of inner disk illumination effects making the NW side brighter than the SE side. %We therefore conclude that the presence of a
Nonetheless, a double ring on its own does not seem able to account for the arm-like structure observed on the NW side of the disk. In Sect.~\ref{sec:flyby} we also showed the hypothesis of a stellar flyby to be very unlikely. We therefore proceed with the evaluation of alternative scenarios. 

\new{Past ALMA sub-millimeter  continuum observations of PDS~70 identified the presence of an overdensity} in the outer disk, near the location of the NIR arm-like structure \new{\cite[Fig.~\ref{fig:ALMAobs};][]{Isella19, ALMAKeppler, Myriam}.
These observations support the hypothesis of the arm-like structure tracing a local rise in density, be it due to a companion-driven spiral density wave (inside the cavity or external to the disk) or to a vortex in the outer disk.}
\new{Considering the proximity between the arm-like structure and planet c, we first explored the hypothesis of a spiral density wave excited by this planet.} %or a vortex in the disk producing a significant overdensity to the NW}.
%\vc{The paragraph above should maybe be the last one of this section (i.e., after the proposed test in the next paragraph)?}
\new{%To investigate whether the feature could be the result of a disk-planet interaction, 
To do so, we monitored the position of the arm-shaped feature over six years in an attempt to find a connection with the orbital motion of planet c, but we did not observe any significant shift of the trace within this time frame.}
Despite the lack of detectable rotation, we cannot exclude the planet-induced spiral hypothesis altogether. It is possible that the rigid-body motion hypothesis is not valid far from the location of the planet causing it (e.g., if the spiral motion slows down at the same time as it gets dissipated as the planet is moving away). This may also explain the lack of (or small amount of) rotation observed for spirals in other protoplanetary disks with a large inner cavity \citep{ren1, ren2, ren3}.

%We also investigated whether the \new{arm-like shaped signal} could trace a spiral arm excited by an external object, such as a flybying star \citep[e.g.,][]{flyby}. More specifically, we explored the possibility that the point source at 2\farcs54 separation from PDS~70 was responsible for such flyby. However, our proper motion and color analysis allowed us to conclude that it is a background star, which cannot dynamically interfere with the disk.

A more likely interpretation for the detected arm-like structure is that it could trace the NIR signature of a vortex, which is trapping dust in a specific part of the outer disk \citep{vortex1, vortex2, vortex3, vortex4}. While the dust overdensity takes the shape of an azimuthal asymmetry in the sub-millimeter range (similar to what is observed with ALMA for PDS~70 in Fig.~\ref{fig:ALMAobs}), it can also appear as a NIR spiral arm for moderate to high inclinations due to flaring and perspective effects \citep{VotexMimicSpiral}. Although a vortex can be indirectly caused by the presence of a planet (through carving of a gap in gas surface density, and subsequent triggering of the Rossby wave instability), it is not expected to follow its motion, and thus would be compatible with our observations. Dedicated hydrodynamical simulations could confirm the vortex hypothesis, as shown in \citet{VotexMimicSpiral} for the case of HD~34282. Another possibility to confirm observationally the vortex hypothesis would be to obtain ALMA gas kinematic observations to probe any deviation from Keplerian motion near the location of the spiral-shaped feature \citep{Garg}. Gas kinematics in a vortex and in planet-induced spirals  indeed carry significantly different signatures \citep{Boehler, Bollati}. 
Monitoring of the arm-shaped structure with additional NIR  observations covering a longer timescale would also enable a better estimation of the angular velocity of the arm-like feature over time, hence allowing us to potentially distinguish whether the structure is static (as in the double-ring disk scenario) or rotating at local Keplerian speed (as expected from the vortex hypothesis).

%Conspicuous \new{spiral arms} have been observed in other disks such as HD~135344B, MWC~758, or CQ~Tau \citep{spiral3,spiral5, Iain}. \new{In particular,} the disk around HD~135344B presents some interesting similarity with PDS~70, since the outer ring asymmetry seen in sub-mm continuum ALMA observations also matches a spiral-arm shape in the near infrared \citep{135344B}.

\new{If the vortex hypothesis is confirmed, one may wonder why no conspicuous planet-driven spiral is observed in the only disk (so far) with two confirmed (multi-Jovian mass) protoplanets. This could raise the  question of whether spiral arms observed in other young disks could be associated with forming planets. The spiral arms observed in the outer disk of other transition disks harboring a large cavity, such as HD~135344B \citep{spiral3} and MWC~758 \citep{spiral5}, could then be caused by physical mechanisms not involving forming planets %on (quasi-) circular orbits 
in the cavity, such as gravitational instability \citep{Lodato2004,Rice2005}, dynamical interactions with an unresolved close binary \citep[e.g.,][]{prince}, %a protoplanet on an eccentric orbit \citep{Calcino2020}, 
a bound companion external to the disk \citep[][]%HD100453
{Dong2016}, or a recent stellar flyby \citep{Menard2020}}. %Hydro-dynamical simulations have indeed shown that multiple spiral arms can result from the interaction of the disc with an inner binary \citep{prince}.
Alternatively, planets on (quasi-)circular orbits such as PDS~70c \citep{Wang21} may cause an outer spiral that is too tightly wound to be distinguishable from the illuminated edge of the outer disk, in particular at large inclinations \citep[][]{Dong2015,Zhu2015}. 
Spirals observed in the outer part of transition disks may nonetheless still be compatible with the presence of embedded planets interior to the spirals, if on significantly eccentric orbits \citep{Calcino2020}.

\section{Conclusion} \label{sec:conclusion}

We designed a new inverse-problem approach-based algorithm (\texttt{mustard}) to reduce ADI data sets while alleviating geometric biases associated with the filtering of azimuthally extended signals, and applied it to reduce all relevant archival PDS~70 data sets obtained by VLT/SPHERE.
Together with all the PDI images available on this source, the new \texttt{mustard} images confirm that an arm-like structure that was tentatively found in previous images is not a geometric bias from ADI-based PSF subtraction methods. It is located radially outward and azimuthally upstream from PDS~70c, and is significantly brighter than disk signals located at a symmetric location with respect to the minor axis.
Monitoring of the trace of the arm-like structure over six years of data does not show any significant motion. %($\gtrsim 0\farcs012$).
This absence of motion is inconsistent with the expectations for a spiral density wave excited by planet c (or any other planet within the cavity)%, which would move by $0\farcs048^{+0\farcs0012}_{-0\farcs003}$
. On the contrary, the structure may correspond to the NIR signature of a vortex, a hypothesis also supported by archival ALMA sub-millimeter continuum data.
Nonetheless, our data cannot rule out the possibility of an asymmetric (or asymmetrically illuminated) double ring. Discarding the latter option would require either follow-up observations to confirm the Keplerian motion of the arm-like structure or high velocity resolution ALMA molecular line observations to probe the kinematics in the vicinity of the structure.

%-----------------------------------------------------------------
\begin{acknowledgements}
This project has received funding from the European Research Council (ERC) under the European Union's Horizon 2020 research and innovation program (grant agreement No 819155), and from the Belgian F.R.S.-FNRS. We are very grateful to Myriam Benisty for her feedback on this work, and to Christophe Pinte, Faustine Cantalloube, Benoit Pairet, Ruobing Dong and R. van Holstein for useful discussions.
This work has made use of data from the European Space Agency (ESA) mission {\it Gaia} (\url{https://www.cosmos.esa.int/gaia}), processed by the {\it Gaia} Data Processing and Analysis Consortium (DPAC, \url{https://www.cosmos.esa.int/web/gaia/dpac/consortium}). Funding for the DPAC has been provided by national institutions, in particular the institutions participating in the {\it Gaia} Multilateral Agreement.
\end{acknowledgements}

\bibliographystyle{aa} % style aa.bst
\bibliography{references}

\begin{thebibliography}{68}
\expandafter\ifx\csname natexlab\endcsname\relax\def\natexlab#1{#1}\fi

\bibitem[{{Adams} \& {Watkins}(1995)}]{vortex3}
{Adams}, F.~C. \& {Watkins}, R. 1995, \apj, 451, 314

\bibitem[{{Bae} \& {Zhu}(2018)}]{Bae}
{Bae}, J. \& {Zhu}, Z. 2018, \apj, 859, 118

\bibitem[{{Bae} {et~al.}(2019){Bae}, {Zhu}, {Baruteau}, {Benisty}, {Dullemond},
  {Facchini}, {Isella}, {Keppler}, {P{\'e}rez}, \& {Teague}}]{Bae19}
{Bae}, J., {Zhu}, Z., {Baruteau}, C., {et~al.} 2019, \apjl, 884, L41

\bibitem[{{Barge} \& {Sommeria}(1995)}]{vortex2}
{Barge}, P. \& {Sommeria}, J. 1995, \aap, 295, L1

\bibitem[{{Benisty} {et~al.}(2021){Benisty}, {Bae}, {Facchini}, {Keppler},
  {Teague}, {Isella}, {Kurtovic}, {P{\'e}rez}, {Sierra}, {Andrews},
  {Carpenter}, {Czekala}, {Dominik}, {Henning}, {Menard}, {Pinilla}, \&
  {Zurlo}}]{Myriam}
{Benisty}, M., {Bae}, J., {Facchini}, S., {et~al.} 2021, \apjl, 916, L2

\bibitem[{{Benisty} {et~al.}(2018){Benisty}, {Juh{\'a}sz}, {Facchini},
  {Pinilla}, {de Boer}, {P{\'e}rez}, {Keppler}, {Muro-Arena}, {Villenave},
  {Andrews}, {Dominik}, {Dullemond}, {Gallenne}, {Garufi}, {Ginski}, \&
  {Isella}}]{HD143006}
{Benisty}, M., {Juh{\'a}sz}, A., {Facchini}, S., {et~al.} 2018, \aap, 619, A171

\bibitem[{{Beuzit} {et~al.}(2019){Beuzit}, {Vigan}, {Mouillet}, {Dohlen},
  {Gratton}, {Boccaletti}, {Sauvage}, {Schmid}, {Langlois}, {Petit},
  {Baruffolo}, {Feldt}, {Milli}, {Wahhaj}, {Abe}, {Anselmi}, {Antichi},
  {Barette}, {Baudrand}, {Baudoz}, {Bazzon}, {Bernardi}, {Blanchard}, {Brast},
  {Bruno}, {Buey}, {Carbillet}, {Carle}, {Cascone}, {Chapron}, {Charton},
  {Chauvin}, {Claudi}, {Costille}, {De Caprio}, {de Boer}, {Delboulb{\'e}},
  {Desidera}, {Dominik}, {Downing}, {Dupuis}, {Fabron}, {Fantinel}, {Farisato},
  {Feautrier}, {Fedrigo}, {Fusco}, {Gigan}, {Ginski}, {Girard}, {Giro},
  {Gisler}, {Gluck}, {Gry}, {Henning}, {Hubin}, {Hugot}, {Incorvaia}, {Jaquet},
  {Kasper}, {Lagadec}, {Lagrange}, {Le Coroller}, {Le Mignant}, {Le Ruyet},
  {Lessio}, {Lizon}, {Llored}, {Lundin}, {Madec}, {Magnard}, {Marteaud},
  {Martinez}, {Maurel}, {M{\'e}nard}, {Mesa}, {M{\"o}ller-Nilsson}, {Moulin},
  {Moutou}, {Orign{\'e}}, {Parisot}, {Pavlov}, {Perret}, {Pragt}, {Puget},
  {Rabou}, {Ramos}, {Reess}, {Rigal}, {Rochat}, {Roelfsema}, {Rousset}, {Roux},
  {Saisse}, {Salasnich}, {Santambrogio}, {Scuderi}, {Segransan}, {Sevin},
  {Siebenmorgen}, {Soenke}, {Stadler}, {Suarez}, {Tiph{\`e}ne}, {Turatto},
  {Udry}, {Vakili}, {Waters}, {Weber}, {Wildi}, {Zins}, \& {Zurlo}}]{sphere}
{Beuzit}, J.~L., {Vigan}, A., {Mouillet}, D., {et~al.} 2019, \aap, 631, A155

\bibitem[{{Boehler} {et~al.}(2021){Boehler}, {M{\'e}nard}, {Robert}, {Isella},
  {Pinte}, {Gonzalez}, {van der Plas}, {Weaver}, {Teague}, {Garg}, \&
  {M{\'e}heut}}]{Boehler}
{Boehler}, Y., {M{\'e}nard}, F., {Robert}, C.~M.~T., {et~al.} 2021, \aap, 650,
  A59

\bibitem[{{Bollati} {et~al.}(2021){Bollati}, {Lodato}, {Price}, \&
  {Pinte}}]{Bollati}
{Bollati}, F., {Lodato}, G., {Price}, D.~J., \& {Pinte}, C. 2021, \mnras, 504,
  5444

\bibitem[{{Calcino} {et~al.}(2020){Calcino}, {Christiaens}, {Price}, {Pinte},
  {Davis}, {van der Marel}, \& {Cuello}}]{Calcino2020}
{Calcino}, J., {Christiaens}, V., {Price}, D.~J., {et~al.} 2020, \mnras, 498,
  639

\bibitem[{{Cantalloube} {et~al.}(2018){Cantalloube}, {Por}, {Dohlen},
  {Sauvage}, {Vigan}, {Kasper}, {Bharmal}, {Henning}, {Brandner}, {Milli},
  {Correia}, \& {Fusco}}]{Faustine2}
{Cantalloube}, F., {Por}, E.~H., {Dohlen}, K., {et~al.} 2018, \aap, 620, L10

\bibitem[{{Christiaens} {et~al.}(2019){Christiaens}, {Casassus}, {Absil},
  {Cantalloube}, {Gomez Gonzalez}, {Girard}, {Ram{\'\i}rez}, {Pairet},
  {Salinas}, {Price}, {Pinte}, {Quanz}, {Jord{\'a}n}, {Mawet}, \&
  {Wahhaj}}]{Valentin}
{Christiaens}, V., {Casassus}, S., {Absil}, O., {et~al.} 2019, \mnras, 486,
  5819

\bibitem[{{Christiaens} {et~al.}(2021){Christiaens}, {Ubeira-Gabellini},
  {C{\'a}novas}, {Delorme}, {Pairet}, {Absil}, {Casassus}, {Girard}, {Zurlo},
  {Aoyama}, {Marleau}, {Spina}, {van der Marel}, {Cieza}, {Lodato},
  {P{\'e}rez}, {Pinte}, {Price}, \& {Reggiani}}]{Valentin2}
{Christiaens}, V., {Ubeira-Gabellini}, M.~G., {C{\'a}novas}, H., {et~al.} 2021,
  \mnras, 502, 6117

\bibitem[{{Cuello} {et~al.}(2020){Cuello}, {Louvet}, {Mentiplay}, {Pinte},
  {Price}, {Winter}, {Nealon}, {M{\'e}nard}, {Lodato}, {Dipierro},
  {Christiaens}, {Montesinos}, {Cuadra}, {Laibe}, {Cieza}, {Dong}, \&
  {Alexander}}]{Cuello2020}
{Cuello}, N., {Louvet}, F., {Mentiplay}, D., {et~al.} 2020, \mnras, 491, 504

\bibitem[{{de Boer} {et~al.}(2021){de Boer}, {Ginski}, {Chauvin}, {M{\'e}nard},
  {Benisty}, {Dominik}, {Maaskant}, {Girard}, {van der Plas}, {Garufi},
  {Perrot}, {Stolker}, {Avenhaus}, {Bohn}, {Delboulb{\'e}}, {Jaquet}, {Buey},
  {M{\"o}ller-Nilsson}, {Pragt}, \& {Fusco}}]{deBoer21}
{de Boer}, J., {Ginski}, C., {Chauvin}, G., {et~al.} 2021, \aap, 649, A25

\bibitem[{{Dong} {et~al.}(2012){Dong}, {Hashimoto}, {Rafikov}, {Zhu},
  {Whitney}, {Kudo}, {Muto}, {Brandt}, {McClure}, {Wisniewski}, {Abe},
  {Brandner}, {Carson}, {Egner}, {Feldt}, {Goto}, {Grady}, {Guyon}, {Hayano},
  {Hayashi}, {Hayashi}, {Henning}, {Hodapp}, {Ishii}, {Iye}, {Janson},
  {Kandori}, {Knapp}, {Kusakabe}, {Kuzuhara}, {Kwon}, {Matsuo}, {McElwain},
  {Miyama}, {Morino}, {Moro-Martin}, {Nishimura}, {Pyo}, {Serabyn}, {Suto},
  {Suzuki}, {Takami}, {Takato}, {Terada}, {Thalmann}, {Tomono}, {Turner},
  {Watanabe}, {Yamada}, {Takami}, {Usuda}, \& {Tamura}}]{Dong12}
{Dong}, R., {Hashimoto}, J., {Rafikov}, R., {et~al.} 2012, \apj, 760, 111

\bibitem[{{Dong} {et~al.}(2016){Dong}, {Zhu}, {Fung}, {Rafikov}, {Chiang}, \&
  {Wagner}}]{Dong2016}
{Dong}, R., {Zhu}, Z., {Fung}, J., {et~al.} 2016, \apjl, 816, L12

\bibitem[{{Dong} {et~al.}(2015){Dong}, {Zhu}, {Rafikov}, \& {Stone}}]{Dong2015}
{Dong}, R., {Zhu}, Z., {Rafikov}, R.~R., \& {Stone}, J.~M. 2015, \apjl, 809, L5

\bibitem[{Flasseur {et~al.}(2021)Flasseur, Thé, Denis, Thiébaut, \&
  Langlois}]{REXPACO}
Flasseur, O., Thé, S., Denis, L., Thiébaut, E., \& Langlois, M. 2021, \aap

\bibitem[{{Gaia Collaboration} {et~al.}(2018){Gaia Collaboration}, {Brown},
  {Vallenari}, {Prusti}, {de Bruijne}, {Babusiaux}, {Bailer-Jones}, {Biermann},
  {Evans}, {Eyer}, {Jansen}, {Jordi}, {Klioner}, {Lammers}, {Lindegren},
  {Luri}, {Mignard}, {Panem}, {Pourbaix}, {Randich}, {Sartoretti}, {Siddiqui},
  {Soubiran}, {van Leeuwen}, {Walton}, {Arenou}, {Bastian}, {Cropper},
  {Drimmel}, {Katz}, {Lattanzi}, {Bakker}, {Cacciari}, {Casta{\~n}eda},
  {Chaoul}, {Cheek}, {De Angeli}, {Fabricius}, {Guerra}, {Holl}, {Masana},
  {Messineo}, {Mowlavi}, {Nienartowicz}, {Panuzzo}, {Portell}, {Riello},
  {Seabroke}, {Tanga}, {Th{\'e}venin}, {Gracia-Abril}, {Comoretto},
  {Garcia-Reinaldos}, {Teyssier}, {Altmann}, {Andrae}, {Audard},
  {Bellas-Velidis}, {Benson}, {Berthier}, {Blomme}, {Burgess}, {Busso},
  {Carry}, {Cellino}, {Clementini}, {Clotet}, {Creevey}, {Davidson}, {De
  Ridder}, {Delchambre}, {Dell'Oro}, {Ducourant},
  {Fern{\'a}ndez-Hern{\'a}ndez}, {Fouesneau}, {Fr{\'e}mat}, {Galluccio},
  {Garc{\'\i}a-Torres}, {Gonz{\'a}lez-N{\'u}{\~n}ez}, {Gonz{\'a}lez-Vidal},
  {Gosset}, {Guy}, {Halbwachs}, {Hambly}, {Harrison}, {Hern{\'a}ndez},
  {Hestroffer}, {Hodgkin}, {Hutton}, {Jasniewicz}, {Jean-Antoine-Piccolo},
  {Jordan}, {Korn}, {Krone-Martins}, {Lanzafame}, {Lebzelter}, {L{\"o}ffler},
  {Manteiga}, {Marrese}, {Mart{\'\i}n-Fleitas}, {Moitinho}, {Mora}, {Muinonen},
  {Osinde}, {Pancino}, {Pauwels}, {Petit}, {Recio-Blanco}, {Richards},
  {Rimoldini}, {Robin}, {Sarro}, {Siopis}, {Smith}, {Sozzetti}, {S{\"u}veges},
  {Torra}, {van Reeven}, {Abbas}, {Abreu Aramburu}, {Accart}, {Aerts},
  {Altavilla}, {{\'A}lvarez}, {Alvarez}, {Alves}, {Anderson}, {Andrei},
  {Anglada Varela}, {Antiche}, {Antoja}, {Arcay}, {Astraatmadja}, {Bach},
  {Baker}, {Balaguer-N{\'u}{\~n}ez}, {Balm}, {Barache}, {Barata}, {Barbato},
  {Barblan}, {Barklem}, {Barrado}, {Barros}, {Barstow}, {Bartholom{\'e}
  Mu{\~n}oz}, {Bassilana}, {Becciani}, {Bellazzini}, {Berihuete}, {Bertone},
  {Bianchi}, {Bienaym{\'e}}, {Blanco-Cuaresma}, {Boch}, {Boeche}, {Bombrun},
  {Borrachero}, {Bossini}, {Bouquillon}, {Bourda}, {Bragaglia}, {Bramante},
  {Breddels}, {Bressan}, {Brouillet}, {Br{\"u}semeister}, {Brugaletta},
  {Bucciarelli}, {Burlacu}, {Busonero}, {Butkevich}, {Buzzi}, {Caffau},
  {Cancelliere}, {Cannizzaro}, {Cantat-Gaudin}, {Carballo}, {Carlucci},
  {Carrasco}, {Casamiquela}, {Castellani}, {Castro-Ginard}, {Charlot},
  {Chemin}, {Chiavassa}, {Cocozza}, {Costigan}, {Cowell}, {Crifo}, {Crosta},
  {Crowley}, {Cuypers}, {Dafonte}, {Damerdji}, {Dapergolas}, {David}, {David},
  {de Laverny}, {De Luise}, {De March}, {de Martino}, {de Souza}, {de Torres},
  {Debosscher}, {del Pozo}, {Delbo}, {Delgado}, {Delgado}, {Di Matteo},
  {Diakite}, {Diener}, {Distefano}, {Dolding}, {Drazinos}, {Dur{\'a}n},
  {Edvardsson}, {Enke}, {Eriksson}, {Esquej}, {Eynard Bontemps}, {Fabre},
  {Fabrizio}, {Faigler}, {Falc{\~a}o}, {Farr{\`a}s Casas}, {Federici},
  {Fedorets}, {Fernique}, {Figueras}, {Filippi}, {Findeisen}, {Fonti},
  {Fraile}, {Fraser}, {Fr{\'e}zouls}, {Gai}, {Galleti}, {Garabato},
  {Garc{\'\i}a-Sedano}, {Garofalo}, {Garralda}, {Gavel}, {Gavras}, {Gerssen},
  {Geyer}, {Giacobbe}, {Gilmore}, {Girona}, {Giuffrida}, {Glass}, {Gomes},
  {Granvik}, {Gueguen}, {Guerrier}, {Guiraud}, {Guti{\'e}rrez-S{\'a}nchez},
  {Haigron}, {Hatzidimitriou}, {Hauser}, {Haywood}, {Heiter}, {Helmi}, {Heu},
  {Hilger}, {Hobbs}, {Hofmann}, {Holland}, {Huckle}, {Hypki}, {Icardi},
  {Jan{\ss}en}, {Jevardat de Fombelle}, {Jonker}, {Juh{\'a}sz}, {Julbe},
  {Karampelas}, {Kewley}, {Klar}, {Kochoska}, {Kohley}, {Kolenberg},
  {Kontizas}, {Kontizas}, {Koposov}, {Kordopatis}, {Kostrzewa-Rutkowska},
  {Koubsky}, {Lambert}, {Lanza}, {Lasne}, {Lavigne}, {Le Fustec}, {Le
  Poncin-Lafitte}, {Lebreton}, {Leccia}, {Leclerc}, {Lecoeur-Taibi},
  {Lenhardt}, {Leroux}, {Liao}, {Licata}, {Lindstr{\o}m}, {Lister}, {Livanou},
  {Lobel}, {L{\'o}pez}, {Managau}, {Mann}, {Mantelet}, {Marchal}, {Marchant},
  {Marconi}, {Marinoni}, {Marschalk{\'o}}, {Marshall}, {Martino}, {Marton},
  {Mary}, {Massari}, {Matijevi{\v{c}}}, {Mazeh}, {McMillan}, {Messina},
  {Michalik}, {Millar}, {Molina}, {Molinaro}, {Moln{\'a}r}, {Montegriffo},
  {Mor}, {Morbidelli}, {Morel}, {Morris}, {Mulone}, {Muraveva}, {Musella},
  {Nelemans}, {Nicastro}, {Noval}, {O'Mullane}, {Ord{\'e}novic},
  {Ord{\'o}{\~n}ez-Blanco}, {Osborne}, {Pagani}, {Pagano}, {Pailler},
  {Palacin}, {Palaversa}, {Panahi}, {Pawlak}, {Piersimoni}, {Pineau}, {Plachy},
  {Plum}, {Poggio}, {Poujoulet}, {Pr{\v{s}}a}, {Pulone}, {Racero}, {Ragaini},
  {Rambaux}, {Ramos-Lerate}, {Regibo}, {Reyl{\'e}}, {Riclet}, {Ripepi}, {Riva},
  {Rivard}, {Rixon}, {Roegiers}, {Roelens}, {Romero-G{\'o}mez}, {Rowell},
  {Royer}, {Ruiz-Dern}, {Sadowski}, {Sagrist{\`a} Sell{\'e}s}, {Sahlmann},
  {Salgado}, {Salguero}, {Sanna}, {Santana-Ros}, {Sarasso}, {Savietto},
  {Schultheis}, {Sciacca}, {Segol}, {Segovia}, {S{\'e}gransan}, {Shih},
  {Siltala}, {Silva}, {Smart}, {Smith}, {Solano}, {Solitro}, {Sordo}, {Soria
  Nieto}, {Souchay}, {Spagna}, {Spoto}, {Stampa}, {Steele},
  {Steidelm{\"u}ller}, {Stephenson}, {Stoev}, {Suess}, {Surdej}, {Szabados},
  {Szegedi-Elek}, {Tapiador}, {Taris}, {Tauran}, {Taylor}, {Teixeira},
  {Terrett}, {Teyssandier}, {Thuillot}, {Titarenko}, {Torra Clotet}, {Turon},
  {Ulla}, {Utrilla}, {Uzzi}, {Vaillant}, {Valentini}, {Valette}, {van Elteren},
  {Van Hemelryck}, {van Leeuwen}, {Vaschetto}, {Vecchiato}, {Veljanoski},
  {Viala}, {Vicente}, {Vogt}, {von Essen}, {Voss}, {Votruba}, {Voutsinas},
  {Walmsley}, {Weiler}, {Wertz}, {Wevers}, {Wyrzykowski}, {Yoldas},
  {{\v{Z}}erjal}, {Ziaeepour}, {Zorec}, {Zschocke}, {Zucker}, {Zurbach}, \&
  {Zwitter}}]{Gaia2}
{Gaia Collaboration}, {Brown}, A.~G.~A., {Vallenari}, A., {et~al.} 2018, \aap,
  616, A1

\bibitem[{{Gaia Collaboration} {et~al.}(2016){Gaia Collaboration}, {Prusti},
  {de Bruijne}, {Brown}, {Vallenari}, {Babusiaux}, {Bailer-Jones}, {Bastian},
  {Biermann}, {Evans}, {Eyer}, {Jansen}, {Jordi}, {Klioner}, {Lammers},
  {Lindegren}, {Luri}, {Mignard}, {Milligan}, {Panem}, {Poinsignon},
  {Pourbaix}, {Randich}, {Sarri}, {Sartoretti}, {Siddiqui}, {Soubiran},
  {Valette}, {van Leeuwen}, {Walton}, {Aerts}, {Arenou}, {Cropper}, {Drimmel},
  {H{\o}g}, {Katz}, {Lattanzi}, {O'Mullane}, {Grebel}, {Holland}, {Huc},
  {Passot}, {Bramante}, {Cacciari}, {Casta{\~n}eda}, {Chaoul}, {Cheek}, {De
  Angeli}, {Fabricius}, {Guerra}, {Hern{\'a}ndez}, {Jean-Antoine-Piccolo},
  {Masana}, {Messineo}, {Mowlavi}, {Nienartowicz}, {Ord{\'o}{\~n}ez-Blanco},
  {Panuzzo}, {Portell}, {Richards}, {Riello}, {Seabroke}, {Tanga},
  {Th{\'e}venin}, {Torra}, {Els}, {Gracia-Abril}, {Comoretto},
  {Garcia-Reinaldos}, {Lock}, {Mercier}, {Altmann}, {Andrae}, {Astraatmadja},
  {Bellas-Velidis}, {Benson}, {Berthier}, {Blomme}, {Busso}, {Carry},
  {Cellino}, {Clementini}, {Cowell}, {Creevey}, {Cuypers}, {Davidson}, {De
  Ridder}, {de Torres}, {Delchambre}, {Dell'Oro}, {Ducourant}, {Fr{\'e}mat},
  {Garc{\'\i}a-Torres}, {Gosset}, {Halbwachs}, {Hambly}, {Harrison}, {Hauser},
  {Hestroffer}, {Hodgkin}, {Huckle}, {Hutton}, {Jasniewicz}, {Jordan},
  {Kontizas}, {Korn}, {Lanzafame}, {Manteiga}, {Moitinho}, {Muinonen},
  {Osinde}, {Pancino}, {Pauwels}, {Petit}, {Recio-Blanco}, {Robin}, {Sarro},
  {Siopis}, {Smith}, {Smith}, {Sozzetti}, {Thuillot}, {van Reeven}, {Viala},
  {Abbas}, {Abreu Aramburu}, {Accart}, {Aguado}, {Allan}, {Allasia},
  {Altavilla}, {{\'A}lvarez}, {Alves}, {Anderson}, {Andrei}, {Anglada Varela},
  {Antiche}, {Antoja}, {Ant{\'o}n}, {Arcay}, {Atzei}, {Ayache}, {Bach},
  {Baker}, {Balaguer-N{\'u}{\~n}ez}, {Barache}, {Barata}, {Barbier}, {Barblan},
  {Baroni}, {Barrado y Navascu{\'e}s}, {Barros}, {Barstow}, {Becciani},
  {Bellazzini}, {Bellei}, {Bello Garc{\'\i}a}, {Belokurov}, {Bendjoya},
  {Berihuete}, {Bianchi}, {Bienaym{\'e}}, {Billebaud}, {Blagorodnova},
  {Blanco-Cuaresma}, {Boch}, {Bombrun}, {Borrachero}, {Bouquillon}, {Bourda},
  {Bouy}, {Bragaglia}, {Breddels}, {Brouillet}, {Br{\"u}semeister},
  {Bucciarelli}, {Budnik}, {Burgess}, {Burgon}, {Burlacu}, {Busonero}, {Buzzi},
  {Caffau}, {Cambras}, {Campbell}, {Cancelliere}, {Cantat-Gaudin}, {Carlucci},
  {Carrasco}, {Castellani}, {Charlot}, {Charnas}, {Charvet}, {Chassat},
  {Chiavassa}, {Clotet}, {Cocozza}, {Collins}, {Collins}, {Costigan}, {Crifo},
  {Cross}, {Crosta}, {Crowley}, {Dafonte}, {Damerdji}, {Dapergolas}, {David},
  {David}, {De Cat}, {de Felice}, {de Laverny}, {De Luise}, {De March}, {de
  Martino}, {de Souza}, {Debosscher}, {del Pozo}, {Delbo}, {Delgado},
  {Delgado}, {di Marco}, {Di Matteo}, {Diakite}, {Distefano}, {Dolding}, {Dos
  Anjos}, {Drazinos}, {Dur{\'a}n}, {Dzigan}, {Ecale}, {Edvardsson}, {Enke},
  {Erdmann}, {Escolar}, {Espina}, {Evans}, {Eynard Bontemps}, {Fabre},
  {Fabrizio}, {Faigler}, {Falc{\~a}o}, {Farr{\`a}s Casas}, {Faye}, {Federici},
  {Fedorets}, {Fern{\'a}ndez-Hern{\'a}ndez}, {Fernique}, {Fienga}, {Figueras},
  {Filippi}, {Findeisen}, {Fonti}, {Fouesneau}, {Fraile}, {Fraser}, {Fuchs},
  {Furnell}, {Gai}, {Galleti}, {Galluccio}, {Garabato}, {Garc{\'\i}a-Sedano},
  {Gar{\'e}}, {Garofalo}, {Garralda}, {Gavras}, {Gerssen}, {Geyer}, {Gilmore},
  {Girona}, {Giuffrida}, {Gomes}, {Gonz{\'a}lez-Marcos},
  {Gonz{\'a}lez-N{\'u}{\~n}ez}, {Gonz{\'a}lez-Vidal}, {Granvik}, {Guerrier},
  {Guillout}, {Guiraud}, {G{\'u}rpide}, {Guti{\'e}rrez-S{\'a}nchez}, {Guy},
  {Haigron}, {Hatzidimitriou}, {Haywood}, {Heiter}, {Helmi}, {Hobbs},
  {Hofmann}, {Holl}, {Holland}, {Hunt}, {Hypki}, {Icardi}, {Irwin}, {Jevardat
  de Fombelle}, {Jofr{\'e}}, {Jonker}, {Jorissen}, {Julbe}, {Karampelas},
  {Kochoska}, {Kohley}, {Kolenberg}, {Kontizas}, {Koposov}, {Kordopatis},
  {Koubsky}, {Kowalczyk}, {Krone-Martins}, {Kudryashova}, {Kull}, {Bachchan},
  {Lacoste-Seris}, {Lanza}, {Lavigne}, {Le Poncin-Lafitte}, {Lebreton},
  {Lebzelter}, {Leccia}, {Leclerc}, {Lecoeur-Taibi}, {Lemaitre}, {Lenhardt},
  {Leroux}, {Liao}, {Licata}, {Lindstr{\o}m}, {Lister}, {Livanou}, {Lobel},
  {L{\"o}ffler}, {L{\'o}pez}, {Lopez-Lozano}, {Lorenz}, {Loureiro},
  {MacDonald}, {Magalh{\~a}es Fernandes}, {Managau}, {Mann}, {Mantelet},
  {Marchal}, {Marchant}, {Marconi}, {Marie}, {Marinoni}, {Marrese},
  {Marschalk{\'o}}, {Marshall}, {Mart{\'\i}n-Fleitas}, {Martino}, {Mary},
  {Matijevi{\v{c}}}, {Mazeh}, {McMillan}, {Messina}, {Mestre}, {Michalik},
  {Millar}, {Miranda}, {Molina}, {Molinaro}, {Molinaro}, {Moln{\'a}r},
  {Moniez}, {Montegriffo}, {Monteiro}, {Mor}, {Mora}, {Morbidelli}, {Morel},
  {Morgenthaler}, {Morley}, {Morris}, {Mulone}, {Muraveva}, {Musella},
  {Narbonne}, {Nelemans}, {Nicastro}, {Noval}, {Ord{\'e}novic},
  {Ordieres-Mer{\'e}}, {Osborne}, {Pagani}, {Pagano}, {Pailler}, {Palacin},
  {Palaversa}, {Parsons}, {Paulsen}, {Pecoraro}, {Pedrosa}, {Pentik{\"a}inen},
  {Pereira}, {Pichon}, {Piersimoni}, {Pineau}, {Plachy}, {Plum}, {Poujoulet},
  {Pr{\v{s}}a}, {Pulone}, {Ragaini}, {Rago}, {Rambaux}, {Ramos-Lerate},
  {Ranalli}, {Rauw}, {Read}, {Regibo}, {Renk}, {Reyl{\'e}}, {Ribeiro},
  {Rimoldini}, {Ripepi}, {Riva}, {Rixon}, {Roelens}, {Romero-G{\'o}mez},
  {Rowell}, {Royer}, {Rudolph}, {Ruiz-Dern}, {Sadowski}, {Sagrist{\`a}
  Sell{\'e}s}, {Sahlmann}, {Salgado}, {Salguero}, {Sarasso}, {Savietto},
  {Schnorhk}, {Schultheis}, {Sciacca}, {Segol}, {Segovia}, {Segransan},
  {Serpell}, {Shih}, {Smareglia}, {Smart}, {Smith}, {Solano}, {Solitro},
  {Sordo}, {Soria Nieto}, {Souchay}, {Spagna}, {Spoto}, {Stampa}, {Steele},
  {Steidelm{\"u}ller}, {Stephenson}, {Stoev}, {Suess}, {S{\"u}veges}, {Surdej},
  {Szabados}, {Szegedi-Elek}, {Tapiador}, {Taris}, {Tauran}, {Taylor},
  {Teixeira}, {Terrett}, {Tingley}, {Trager}, {Turon}, {Ulla}, {Utrilla},
  {Valentini}, {van Elteren}, {Van Hemelryck}, {van Leeuwen}, {Varadi},
  {Vecchiato}, {Veljanoski}, {Via}, {Vicente}, {Vogt}, {Voss}, {Votruba},
  {Voutsinas}, {Walmsley}, {Weiler}, {Weingrill}, {Werner}, {Wevers},
  {Whitehead}, {Wyrzykowski}, {Yoldas}, {{\v{Z}}erjal}, {Zucker}, {Zurbach},
  {Zwitter}, {Alecu}, {Allen}, {Allende Prieto}, {Amorim},
  {Anglada-Escud{\'e}}, {Arsenijevic}, {Azaz}, {Balm}, {Beck}, {Bernstein},
  {Bigot}, {Bijaoui}, {Blasco}, {Bonfigli}, {Bono}, {Boudreault}, {Bressan},
  {Brown}, {Brunet}, {Bunclark}, {Buonanno}, {Butkevich}, {Carret}, {Carrion},
  {Chemin}, {Ch{\'e}reau}, {Corcione}, {Darmigny}, {de Boer}, {de Teodoro}, {de
  Zeeuw}, {Delle Luche}, {Domingues}, {Dubath}, {Fodor}, {Fr{\'e}zouls},
  {Fries}, {Fustes}, {Fyfe}, {Gallardo}, {Gallegos}, {Gardiol}, {Gebran},
  {Gomboc}, {G{\'o}mez}, {Grux}, {Gueguen}, {Heyrovsky}, {Hoar}, {Iannicola},
  {Isasi Parache}, {Janotto}, {Joliet}, {Jonckheere}, {Keil}, {Kim},
  {Klagyivik}, {Klar}, {Knude}, {Kochukhov}, {Kolka}, {Kos}, {Kutka}, {Lainey},
  {LeBouquin}, {Liu}, {Loreggia}, {Makarov}, {Marseille}, {Martayan},
  {Martinez-Rubi}, {Massart}, {Meynadier}, {Mignot}, {Munari}, {Nguyen},
  {Nordlander}, {Ocvirk}, {O'Flaherty}, {Olias Sanz}, {Ortiz}, {Osorio},
  {Oszkiewicz}, {Ouzounis}, {Palmer}, {Park}, {Pasquato}, {Peltzer}, {Peralta},
  {P{\'e}turaud}, {Pieniluoma}, {Pigozzi}, {Poels}, {Prat}, {Prod'homme},
  {Raison}, {Rebordao}, {Risquez}, {Rocca-Volmerange}, {Rosen}, {Ruiz-Fuertes},
  {Russo}, {Sembay}, {Serraller Vizcaino}, {Short}, {Siebert}, {Silva},
  {Sinachopoulos}, {Slezak}, {Soffel}, {Sosnowska}, {Strai{\v{z}}ys}, {ter
  Linden}, {Terrell}, {Theil}, {Tiede}, {Troisi}, {Tsalmantza}, {Tur},
  {Vaccari}, {Vachier}, {Valles}, {Van Hamme}, {Veltz}, {Virtanen}, {Wallut},
  {Wichmann}, {Wilkinson}, {Ziaeepour}, \& {Zschocke}}]{Gaia}
{Gaia Collaboration}, {Prusti}, T., {de Bruijne}, J.~H.~J., {et~al.} 2016,
  \aap, 595, A1

\bibitem[{{Garg} {et~al.}(2021){Garg}, {Pinte}, {Christiaens}, {Price},
  {Lazendic}, {Boehler}, {Casassus}, {Marino}, {Perez}, \& {Zuleta}}]{Garg}
{Garg}, H., {Pinte}, C., {Christiaens}, V., {et~al.} 2021, \mnras, 504, 782

\bibitem[{{Goldreich} \& {Tremaine}(1979)}]{Lindblad}
{Goldreich}, P. \& {Tremaine}, S. 1979, \apj, 233, 857

\bibitem[{{Gomez Gonzalez} {et~al.}(2017){Gomez Gonzalez}, {Wertz}, {Absil},
  {Christiaens}, {Defr{\`e}re}, {Mawet}, {Milli}, {Absil}, {Van Droogenbroeck},
  {Cantalloube}, {Hinz}, {Skemer}, {Karlsson}, \& {Surdej}}]{vipref}
{Gomez Gonzalez}, C.~A., {Wertz}, O., {Absil}, O., {et~al.} 2017, \aj, 154, 7

\bibitem[{{Grady} {et~al.}(2013){Grady}, {Muto}, {Hashimoto}, {Fukagawa},
  {Currie}, {Biller}, {Thalmann}, {Sitko}, {Russell}, {Wisniewski}, {Dong},
  {Kwon}, {Sai}, {Hornbeck}, {Schneider}, {Hines}, {Moro Mart{\'\i}n}, {Feldt},
  {Henning}, {Pott}, {Bonnefoy}, {Bouwman}, {Lacour}, {Mueller}, {Juh{\'a}sz},
  {Crida}, {Chauvin}, {Andrews}, {Wilner}, {Kraus}, {Dahm}, {Robitaille},
  {Jang-Condell}, {Abe}, {Akiyama}, {Brandner}, {Brandt}, {Carson}, {Egner},
  {Follette}, {Goto}, {Guyon}, {Hayano}, {Hayashi}, {Hayashi}, {Hodapp},
  {Ishii}, {Iye}, {Janson}, {Kandori}, {Knapp}, {Kudo}, {Kusakabe}, {Kuzuhara},
  {Mayama}, {McElwain}, {Matsuo}, {Miyama}, {Morino}, {Nishimura}, {Pyo},
  {Serabyn}, {Suto}, {Suzuki}, {Takami}, {Takato}, {Terada}, {Tomono},
  {Turner}, {Watanabe}, {Yamada}, {Takami}, {Usuda}, \& {Tamura}}]{spiral5}
{Grady}, C.~A., {Muto}, T., {Hashimoto}, J., {et~al.} 2013, \apj, 762, 48

\bibitem[{{Haffert} {et~al.}(2019){Haffert}, {Bohn}, {de Boer}, {Snellen},
  {Brinchmann}, {Girard}, {Keller}, \& {Bacon}}]{PDS_first_claim_planet_c}
{Haffert}, S.~Y., {Bohn}, A.~J., {de Boer}, J., {et~al.} 2019, Nature
  Astronomy, 3, 749

\bibitem[{{Hashimoto} {et~al.}(2012){Hashimoto}, {Dong}, {Kudo}, {Honda},
  {McClure}, {Zhu}, {Muto}, {Wisniewski}, {Abe}, {Brandner}, {Brandt},
  {Carson}, {Egner}, {Feldt}, {Fukagawa}, {Goto}, {Grady}, {Guyon}, {Hayano},
  {Hayashi}, {Hayashi}, {Henning}, {Hodapp}, {Ishii}, {Iye}, {Janson},
  {Kandori}, {Knapp}, {Kusakabe}, {Kuzuhara}, {Kwon}, {Matsuo}, {Mayama},
  {McElwain}, {Miyama}, {Morino}, {Moro-Martin}, {Nishimura}, {Pyo}, {Serabyn},
  {Suenaga}, {Suto}, {Suzuki}, {Takahashi}, {Takami}, {Takato}, {Terada},
  {Thalmann}, {Tomono}, {Turner}, {Watanabe}, {Yamada}, {Takami}, {Usuda}, \&
  {Tamura}}]{Hashimoto}
{Hashimoto}, J., {Dong}, R., {Kudo}, T., {et~al.} 2012, \apjl, 758, L19

\bibitem[{{Isella} {et~al.}(2019){Isella}, {Benisty}, {Teague}, {Bae},
  {Keppler}, {Facchini}, \& {P{\'e}rez}}]{Isella19}
{Isella}, A., {Benisty}, M., {Teague}, R., {et~al.} 2019, \apjl, 879, L25

\bibitem[{{Keppler} {et~al.}(2018){Keppler}, {Benisty}, {M{\"u}ller},
  {Henning}, {van Boekel}, {Cantalloube}, {Ginski}, {van Holstein}, {Maire},
  {Pohl}, {Samland}, {Avenhaus}, {Baudino}, {Boccaletti}, {de Boer},
  {Bonnefoy}, {Chauvin}, {Desidera}, {Langlois}, {Lazzoni}, {Marleau},
  {Mordasini}, {Pawellek}, {Stolker}, {Vigan}, {Zurlo}, {Birnstiel},
  {Brandner}, {Feldt}, {Flock}, {Girard}, {Gratton}, {Hagelberg}, {Isella},
  {Janson}, {Juhasz}, {Kemmer}, {Kral}, {Lagrange}, {Launhardt}, {Matter},
  {M{\'e}nard}, {Milli}, {Molli{\`e}re}, {Olofsson}, {P{\'e}rez}, {Pinilla},
  {Pinte}, {Quanz}, {Schmidt}, {Udry}, {Wahhaj}, {Williams}, {Buenzli},
  {Cudel}, {Dominik}, {Galicher}, {Kasper}, {Lannier}, {Mesa}, {Mouillet},
  {Peretti}, {Perrot}, {Salter}, {Sissa}, {Wildi}, {Abe}, {Antichi},
  {Augereau}, {Baruffolo}, {Baudoz}, {Bazzon}, {Beuzit}, {Blanchard}, {Brems},
  {Buey}, {De Caprio}, {Carbillet}, {Carle}, {Cascone}, {Cheetham}, {Claudi},
  {Costille}, {Delboulb{\'e}}, {Dohlen}, {Fantinel}, {Feautrier}, {Fusco},
  {Giro}, {Gluck}, {Gry}, {Hubin}, {Hugot}, {Jaquet}, {Le Mignant}, {Llored},
  {Madec}, {Magnard}, {Martinez}, {Maurel}, {Meyer}, {M{\"o}ller-Nilsson},
  {Moulin}, {Mugnier}, {Orign{\'e}}, {Pavlov}, {Perret}, {Petit}, {Pragt},
  {Puget}, {Rabou}, {Ramos}, {Rigal}, {Rochat}, {Roelfsema}, {Rousset}, {Roux},
  {Salasnich}, {Sauvage}, {Sevin}, {Soenke}, {Stadler}, {Suarez}, {Turatto}, \&
  {Weber}}]{PDS70info}
{Keppler}, M., {Benisty}, M., {M{\"u}ller}, A., {et~al.} 2018, \aap, 617, A44

\bibitem[{{Keppler} {et~al.}(2019){Keppler}, {Teague}, {Bae}, {Benisty},
  {Henning}, {van Boekel}, {Chapillon}, {Pinilla}, {Williams}, {Bertrang},
  {Facchini}, {Flock}, {Ginski}, {Juhasz}, {Klahr}, {Liu}, {M{\"u}ller},
  {P{\'e}rez}, {Pohl}, {Rosotti}, {Samland}, \& {Semenov}}]{ALMAKeppler}
{Keppler}, M., {Teague}, R., {Bae}, J., {et~al.} 2019, \aap, 625, A118

\bibitem[{{Lenzen} {et~al.}(2003){Lenzen}, {Hartung}, {Brandner}, {Finger},
  {Hubin}, {Lacombe}, {Lagrange}, {Lehnert}, {Moorwood}, \&
  {Mouillet}}]{Lenzen}
{Lenzen}, R., {Hartung}, M., {Brandner}, W., {et~al.} 2003, in Society of
  Photo-Optical Instrumentation Engineers (SPIE) Conference Series, Vol. 4841,
  Instrument Design and Performance for Optical/Infrared Ground-based
  Telescopes, ed. M.~{Iye} \& A.~F.~M. {Moorwood}, 944--952

\bibitem[{{Lodato} \& {Rice}(2004)}]{Lodato2004}
{Lodato}, G. \& {Rice}, W.~K.~M. 2004, \mnras, 351, 630

\bibitem[{{Long} {et~al.}(2018){Long}, {Akiyama}, {Sitko}, {Fernandes},
  {Assani}, {Grady}, {Cure}, {Danchi}, {Dong}, {Fukagawa}, {Hasegawa},
  {Hashimoto}, {Henning}, {Inutsuka}, {Kraus}, {Kwon}, {Lisse}, {Baobabu Liu},
  {Mayama}, {Muto}, {Nakagawa}, {Takami}, {Tamura}, {Currie}, {Wisniewski}, \&
  {Yang}}]{Long}
{Long}, Z.~C., {Akiyama}, E., {Sitko}, M., {et~al.} 2018, \apj, 858, 112

\bibitem[{{Lovelace} {et~al.}(1999){Lovelace}, {Li}, {Colgate}, \&
  {Nelson}}]{vortex_planet}
{Lovelace}, R.~V.~E., {Li}, H., {Colgate}, S.~A., \& {Nelson}, A.~F. 1999,
  \apj, 513, 805

\bibitem[{{Lyra} \& {Lin}(2013)}]{vortex1}
{Lyra}, W. \& {Lin}, M.-K. 2013, \apj, 775, 17

\bibitem[{{Ma} {et~al.}(2022){Ma}, {De Rosa}, \& {Kalas}}]{flyby}
{Ma}, Y., {De Rosa}, R.~J., \& {Kalas}, P. 2022, \aj, 163, 219

\bibitem[{Marr \& Dong(2022)}]{VotexMimicSpiral}
Marr, M. \& Dong, R. 2022, The Astrophysical Journal, 930, 80

\bibitem[{{M{\'e}nard} {et~al.}(2020){M{\'e}nard}, {Cuello}, {Ginski}, {van der
  Plas}, {Villenave}, {Gonzalez}, {Pinte}, {Benisty}, {Boccaletti}, {Price},
  {Boehler}, {Chripko}, {de Boer}, {Dominik}, {Garufi}, {Gratton}, {Hagelberg},
  {Henning}, {Langlois}, {Maire}, {Pinilla}, {Ruane}, {Schmid}, {van Holstein},
  {Vigan}, {Zurlo}, {Hubin}, {Pavlov}, {Rochat}, {Sauvage}, \&
  {Stadler}}]{Menard2020}
{M{\'e}nard}, F., {Cuello}, N., {Ginski}, C., {et~al.} 2020, \aap, 639, L1

\bibitem[{{Mesa} {et~al.}(2019){Mesa}, {Keppler}, {Cantalloube}, {Rodet},
  {Charnay}, {Gratton}, {Langlois}, {Boccaletti}, {Bonnefoy}, {Vigan},
  {Flasseur}, {Bae}, {Benisty}, {Chauvin}, {de Boer}, {Desidera}, {Henning},
  {Lagrange}, {Meyer}, {Milli}, {M{\"u}ller}, {Pairet}, {Zurlo}, {Antoniucci},
  {Baudino}, {Brown Sevilla}, {Cascone}, {Cheetham}, {Claudi}, {Delorme},
  {D'Orazi}, {Feldt}, {Hagelberg}, {Janson}, {Kral}, {Lagadec}, {Lazzoni},
  {Ligi}, {Maire}, {Martinez}, {Menard}, {Meunier}, {Perrot}, {Petrus},
  {Pinte}, {Rickman}, {Rochat}, {Rouan}, {Samland}, {Sauvage}, {Schmidt},
  {Udry}, {Weber}, \& {Wildi}}]{periodPlanetc}
{Mesa}, D., {Keppler}, M., {Cantalloube}, F., {et~al.} 2019, \aap, 632, A25

\bibitem[{{Milli} {et~al.}(2012){Milli}, {Mouillet}, {Lagrange}, {Boccaletti},
  {Mawet}, {Chauvin}, \& {Bonnefoy}}]{Milli}
{Milli}, J., {Mouillet}, D., {Lagrange}, A.~M., {et~al.} 2012, \aap, 545, A111

\bibitem[{{Montesinos} {et~al.}(2016){Montesinos}, {Perez}, {Casassus},
  {Marino}, {Cuadra}, \& {Christiaens}}]{Montesinos16}
{Montesinos}, M., {Perez}, S., {Casassus}, S., {et~al.} 2016, \apjl, 823, L8

\bibitem[{{Mouillet} {et~al.}(2001){Mouillet}, {Lagrange}, {Augereau}, \&
  {M{\'e}nard}}]{spiral2}
{Mouillet}, D., {Lagrange}, A.~M., {Augereau}, J.~C., \& {M{\'e}nard}, F. 2001,
  \aap, 372, L61

\bibitem[{{M{\"u}ller} {et~al.}(2018){M{\"u}ller}, {Keppler}, {Henning},
  {Samland}, {Chauvin}, {Beust}, {Maire}, {Molaverdikhani}, {van Boekel},
  {Benisty}, {Boccaletti}, {Bonnefoy}, {Cantalloube}, {Charnay}, {Baudino},
  {Gennaro}, {Long}, {Cheetham}, {Desidera}, {Feldt}, {Fusco}, {Girard},
  {Gratton}, {Hagelberg}, {Janson}, {Lagrange}, {Langlois}, {Lazzoni}, {Ligi},
  {M{\'e}nard}, {Mesa}, {Meyer}, {Molli{\`e}re}, {Mordasini}, {Moulin},
  {Pavlov}, {Pawellek}, {Quanz}, {Ramos}, {Rouan}, {Sissa}, {Stadler}, {Vigan},
  {Wahhaj}, {Weber}, \& {Zurlo}}]{muller}
{M{\"u}ller}, A., {Keppler}, M., {Henning}, T., {et~al.} 2018, \aap, 617, L2

\bibitem[{{Muto} {et~al.}(2012){Muto}, {Grady}, {Hashimoto}, {Fukagawa},
  {Hornbeck}, {Sitko}, {Russell}, {Werren}, {Cur{\'e}}, {Currie}, {Ohashi},
  {Okamoto}, {Momose}, {Honda}, {Inutsuka}, {Takeuchi}, {Dong}, {Abe},
  {Brandner}, {Brandt}, {Carson}, {Egner}, {Feldt}, {Fukue}, {Goto}, {Guyon},
  {Hayano}, {Hayashi}, {Hayashi}, {Henning}, {Hodapp}, {Ishii}, {Iye},
  {Janson}, {Kandori}, {Knapp}, {Kudo}, {Kusakabe}, {Kuzuhara}, {Matsuo},
  {Mayama}, {McElwain}, {Miyama}, {Morino}, {Moro-Martin}, {Nishimura}, {Pyo},
  {Serabyn}, {Suto}, {Suzuki}, {Takami}, {Takato}, {Terada}, {Thalmann},
  {Tomono}, {Turner}, {Watanabe}, {Wisniewski}, {Yamada}, {Takami}, {Usuda}, \&
  {Tamura}}]{spiral3}
{Muto}, T., {Grady}, C.~A., {Hashimoto}, J., {et~al.} 2012, \apjl, 748, L22

\bibitem[{{Nelson} {et~al.}(1998){Nelson}, {Benz}, {Adams}, \&
  {Arnett}}]{selfgravity}
{Nelson}, A.~F., {Benz}, W., {Adams}, F.~C., \& {Arnett}, D. 1998, \apj, 502,
  342

\bibitem[{{Owen} \& {Kollmeier}(2017)}]{vortex_formation}
{Owen}, J.~E. \& {Kollmeier}, J.~A. 2017, \mnras, 467, 3379

\bibitem[{{Pairet} {et~al.}(2021){Pairet}, {Cantalloube}, \& {Jacques}}]{mayo}
{Pairet}, B., {Cantalloube}, F., \& {Jacques}, L. 2021, \mnras, 503, 3724

\bibitem[{{Pantin} {et~al.}(2000){Pantin}, {Waelkens}, \& {Lagage}}]{spiral1}
{Pantin}, E., {Waelkens}, C., \& {Lagage}, P.~O. 2000, \aap, 361, L9

\bibitem[{{Pinte} {et~al.}(2009){Pinte}, {Harries}, {Min}, {Watson},
  {Dullemond}, {Woitke}, {M{\'e}nard}, \& {Dur{\'a}n-Rojas}}]{Pinte2009}
{Pinte}, C., {Harries}, T.~J., {Min}, M., {et~al.} 2009, \aap, 498, 967

\bibitem[{{Pinte} {et~al.}(2006){Pinte}, {M{\'e}nard}, {Duch{\^e}ne}, \&
  {Bastien}}]{MCFOST}
{Pinte}, C., {M{\'e}nard}, F., {Duch{\^e}ne}, G., \& {Bastien}, P. 2006, \aap,
  459, 797

\bibitem[{{Price} {et~al.}(2018){Price}, {Cuello}, {Pinte}, {Mentiplay},
  {Casassus}, {Christiaens}, {Kennedy}, {Cuadra}, {Sebastian Perez}, {Marino},
  {Armitage}, {Zurlo}, {Juhasz}, {Ragusa}, {Laibe}, \& {Lodato}}]{prince}
{Price}, D.~J., {Cuello}, N., {Pinte}, C., {et~al.} 2018, \mnras, 477, 1270

\bibitem[{{Quillen}(2006)}]{shadowstudy}
{Quillen}, A.~C. 2006, \apj, 640, 1078

\bibitem[{{Rafikov}(2002)}]{Rafikov}
{Rafikov}, R.~R. 2002, \apj, 569, 997

\bibitem[{{Reg{\'a}ly} {et~al.}(2021){Reg{\'a}ly}, {Kadam}, \&
  {Dullemond}}]{vortex_no_planet}
{Reg{\'a}ly}, Z., {Kadam}, K., \& {Dullemond}, C.~P. 2021, \mnras, 506, 2685

\bibitem[{{Reggiani} {et~al.}(2018){Reggiani}, {Christiaens}, {Absil}, {Mawet},
  {Huby}, {Choquet}, {Gomez Gonzalez}, {Ruane}, {Femenia}, {Serabyn},
  {Matthews}, {Barraza}, {Carlomagno}, {Defr{\`e}re}, {Delacroix}, {Habraken},
  {Jolivet}, {Karlsson}, {Orban de Xivry}, {Piron}, {Surdej}, {Vargas Catalan},
  \& {Wertz}}]{MWC_wavelenght_dependecies}
{Reggiani}, M., {Christiaens}, V., {Absil}, O., {et~al.} 2018, \aap, 611, A74

\bibitem[{{Ren} {et~al.}(2018){Ren}, {Dong}, {Esposito}, {Pueyo}, {Debes},
  {Poteet}, {Choquet}, {Benisty}, {Chiang}, {Grady}, {Hines}, {Schneider}, \&
  {Soummer}}]{ren1}
{Ren}, B., {Dong}, R., {Esposito}, T.~M., {et~al.} 2018, \apjl, 857, L9

\bibitem[{{Ren} {et~al.}(2020){Ren}, {Dong}, {van Holstein}, {Ruffio},
  {Calvin}, {Girard}, {Benisty}, {Boccaletti}, {Esposito}, {Choquet}, {Mawet},
  {Pueyo}, {Stolker}, {Chiang}, {Boer}, {Debes}, {Garufi}, {Grady}, {Hines},
  {Maire}, {M{\'e}nard}, {Millar-Blanchaer}, {Perrin}, {Poteet}, \&
  {Schneider}}]{ren2}
{Ren}, B., {Dong}, R., {van Holstein}, R.~G., {et~al.} 2020, \apjl, 898, L38

\bibitem[{{Riaud} {et~al.}(2006){Riaud}, {Mawet}, {Absil}, {Boccaletti},
  {Baudoz}, {Herwats}, \& {Surdej}}]{NACO_PDS}
{Riaud}, P., {Mawet}, D., {Absil}, O., {et~al.} 2006, \aap, 458, 317

\bibitem[{{Rice} {et~al.}(2005){Rice}, {Lodato}, \& {Armitage}}]{Rice2005}
{Rice}, W.~K.~M., {Lodato}, G., \& {Armitage}, P.~J. 2005, \mnras, 364, L56

\bibitem[{{Rousset} {et~al.}(2003){Rousset}, {Lacombe}, {Puget}, {Hubin},
  {Gendron}, {Fusco}, {Arsenault}, {Charton}, {Feautrier}, {Gigan}, {Kern},
  {Lagrange}, {Madec}, {Mouillet}, {Rabaud}, {Rabou}, {Stadler}, \&
  {Zins}}]{Rousset}
{Rousset}, G., {Lacombe}, F., {Puget}, P., {et~al.} 2003, in Society of
  Photo-Optical Instrumentation Engineers (SPIE) Conference Series, Vol. 4839,
  Adaptive Optical System Technologies II, ed. P.~L. {Wizinowich} \&
  D.~{Bonaccini}, 140--149

\bibitem[{{Stolker} {et~al.}(2016){Stolker}, {Dominik}, {Min}, {Garufi},
  {Mulders}, \& {Avenhaus}}]{diskmap}
{Stolker}, T., {Dominik}, C., {Min}, M., {et~al.} 2016, \aap, 596, A70

\bibitem[{{Tanga} {et~al.}(1996){Tanga}, {Babiano}, {Dubrulle}, \&
  {Provenzale}}]{vortex4}
{Tanga}, P., {Babiano}, A., {Dubrulle}, B., \& {Provenzale}, A. 1996, \icarus,
  121, 158

\bibitem[{{Toci} {et~al.}(2020){Toci}, {Lodato}, {Christiaens}, {Fedele},
  {Pinte}, {Price}, \& {Testi}}]{Toci20}
{Toci}, C., {Lodato}, G., {Christiaens}, V., {et~al.} 2020, \mnras, 499, 2015

\bibitem[{{van Holstein} {et~al.}(2020){van Holstein}, {Girard}, {de Boer},
  {Snik}, {Milli}, {Stam}, {Ginski}, {Mouillet}, {Wahhaj}, {Schmid}, {Keller},
  {Langlois}, {Dohlen}, {Vigan}, {Pohl}, {Carbillet}, {Fantinel}, {Maurel},
  {Orign{\'e}}, {Petit}, {Ramos}, {Rigal}, {Sevin}, {Boccaletti}, {Le
  Coroller}, {Dominik}, {Henning}, {Lagadec}, {M{\'e}nard}, {Turatto}, {Udry},
  {Chauvin}, {Feldt}, \& {Beuzit}}]{irdap}
{van Holstein}, R.~G., {Girard}, J.~H., {de Boer}, J., {et~al.} 2020, \aap,
  633, A64

\bibitem[{{van Holstein} {et~al.}(2021){van Holstein}, {Stolker},
  {Jensen-Clem}, {Ginski}, {Milli}, {de Boer}, {Girard}, {Wahhaj}, {Bohn},
  {Millar-Blanchaer}, {Benisty}, {Bonnefoy}, {Chauvin}, {Dominik}, {Hinkley},
  {Keller}, {Keppler}, {Langlois}, {Marino}, {M{\'e}nard}, {Perrot}, {Schmidt},
  {Vigan}, {Zurlo}, \& {Snik}}]{Holstein21}
{van Holstein}, R.~G., {Stolker}, T., {Jensen-Clem}, R., {et~al.} 2021, \aap,
  647, A21

\bibitem[{{Wang} {et~al.}(2021){Wang}, {Vigan}, {Lacour}, {Nowak}, {Stolker},
  {De Rosa}, {Ginzburg}, {Gao}, {Abuter}, {Amorim}, {Asensio-Torres},
  {Baub{\"o}ck}, {Benisty}, {Berger}, {Beust}, {Beuzit}, {Blunt}, {Boccaletti},
  {Bohn}, {Bonnefoy}, {Bonnet}, {Brandner}, {Cantalloube}, {Caselli},
  {Charnay}, {Chauvin}, {Choquet}, {Christiaens}, {Cl{\'e}net}, {Coud{\'e} Du
  Foresto}, {Cridland}, {de Zeeuw}, {Dembet}, {Dexter}, {Drescher}, {Duvert},
  {Eckart}, {Eisenhauer}, {Facchini}, {Gao}, {Garcia}, {Garcia Lopez},
  {Gardner}, {Gendron}, {Genzel}, {Gillessen}, {Girard}, {Haubois},
  {Hei{\ss}el}, {Henning}, {Hinkley}, {Hippler}, {Horrobin}, {Houll{\'e}},
  {Hubert}, {Jim{\'e}nez-Rosales}, {Jocou}, {Kammerer}, {Keppler}, {Kervella},
  {Meyer}, {Kreidberg}, {Lagrange}, {Lapeyr{\`e}re}, {Le Bouquin}, {L{\'e}na},
  {Lutz}, {Maire}, {M{\'e}nard}, {M{\'e}rand}, {Molli{\`e}re}, {Monnier},
  {Mouillet}, {M{\"u}ller}, {Nasedkin}, {Ott}, {Otten}, {Paladini}, {Paumard},
  {Perraut}, {Perrin}, {Pfuhl}, {Pueyo}, {Rameau}, {Rodet},
  {Rodr{\'\i}guez-Coira}, {Rousset}, {Scheithauer}, {Shangguan}, {Shimizu},
  {Stadler}, {Straub}, {Straubmeier}, {Sturm}, {Tacconi}, {van Dishoeck},
  {Vincent}, {von Fellenberg}, {Ward-Duong}, {Widmann}, {Wieprecht},
  {Wiezorrek}, {Woillez}, \& {Gravity Collaboration}}]{Wang21}
{Wang}, J.~J., {Vigan}, A., {Lacour}, S., {et~al.} 2021, \aj, 161, 148

\bibitem[{{Xie} {et~al.}(2021){Xie}, {Ren}, {Dong}, {Pueyo}, {Ruffio}, {Fang},
  {Mawet}, \& {Stolker}}]{ren3}
{Xie}, C., {Ren}, B., {Dong}, R., {et~al.} 2021, \apjl, 906, L9

\bibitem[{{Zhu} {et~al.}(2015){Zhu}, {Dong}, {Stone}, \& {Rafikov}}]{Zhu2015}
{Zhu}, Z., {Dong}, R., {Stone}, J.~M., \& {Rafikov}, R.~R. 2015, \apj, 813, 88

\end{thebibliography}

\begin{appendix}
\begin{figure*}[!t]
    \section{Gallery of reductions}
    \label{sec:gallery}
    \centering
    \includegraphics[width=\linewidth]{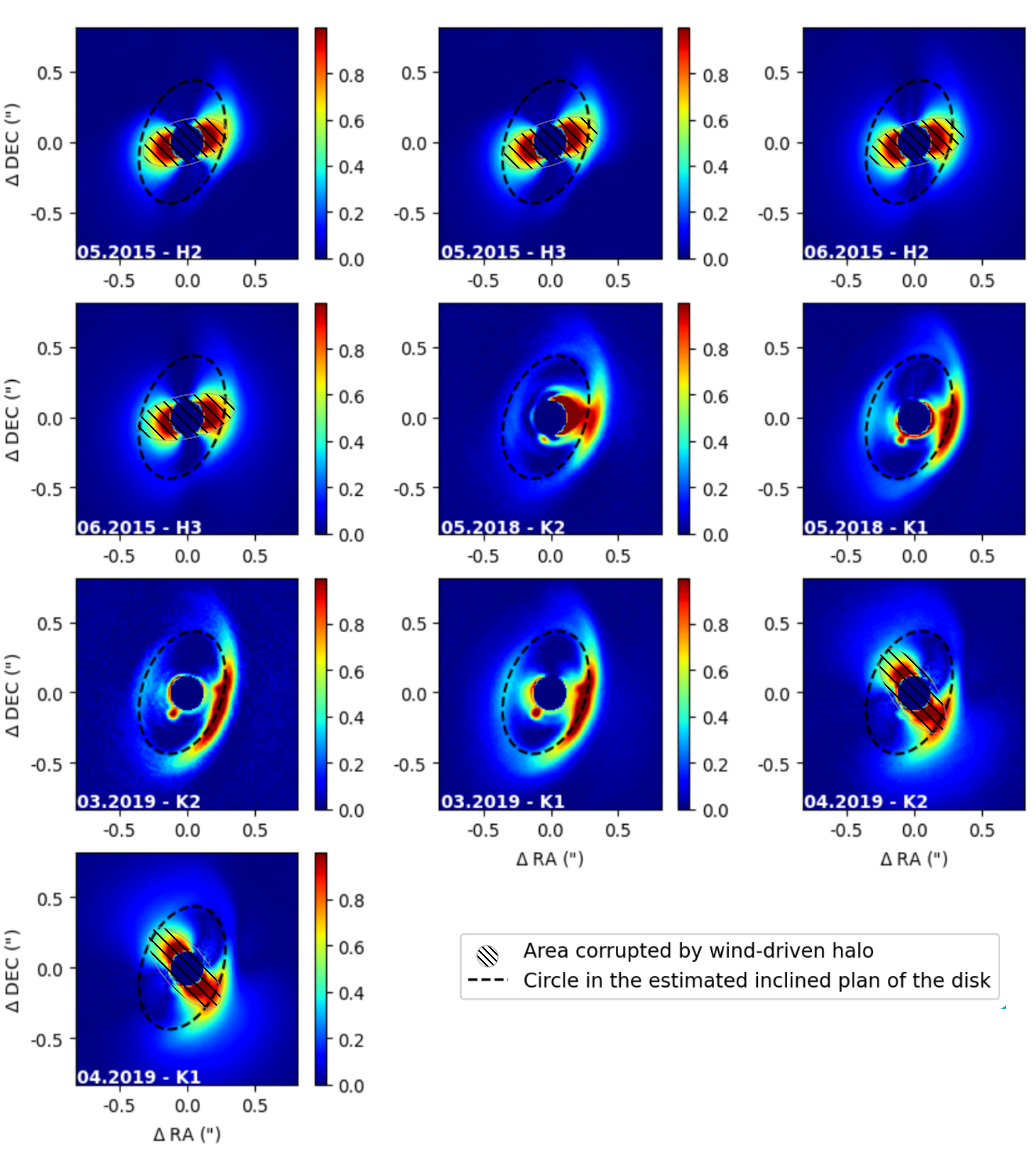}
     \caption{Gallery of reductions of the ADI data sets provided by the \texttt{mustard} algorithm (see Sect.~\ref{sec:reduction} for details). Each data set is obtained in two separate filters (K1/K2 or H2/H3). Images are normalized by their maximum values. }
     \label{fig:ADIreduced}
\end{figure*}

\begin{figure*}[!t]
    \centering
    \includegraphics[width=\linewidth]{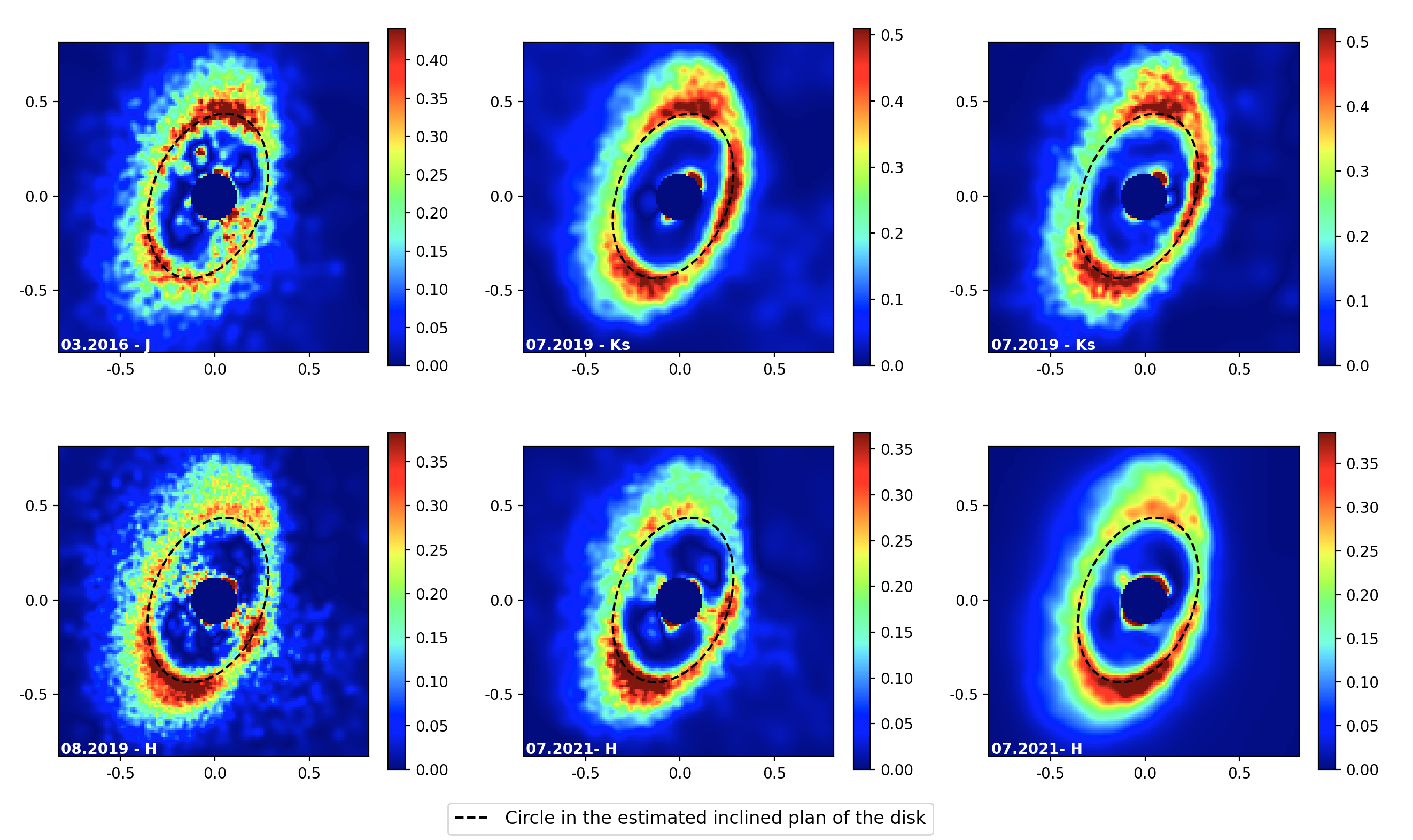}
     \caption{Gallery of reductions obtained for the PDI data sets, showing the ${Q_\phi}^2$ images provided by the \texttt{irdap} pipeline (see Sect.~\ref{sec:reduction} for details). Two PDI data sets were split into two subsets, due to different coronagraph settings. Images are normalized by their maximum values. }
     \label{fig:PDIreduced}
\end{figure*}

\clearpage
\section{Effect of the Laplacian filter on the images}
\label{sec:laplace}
\begin{figure}%[H]
    \centering
    \includegraphics[width=\linewidth]{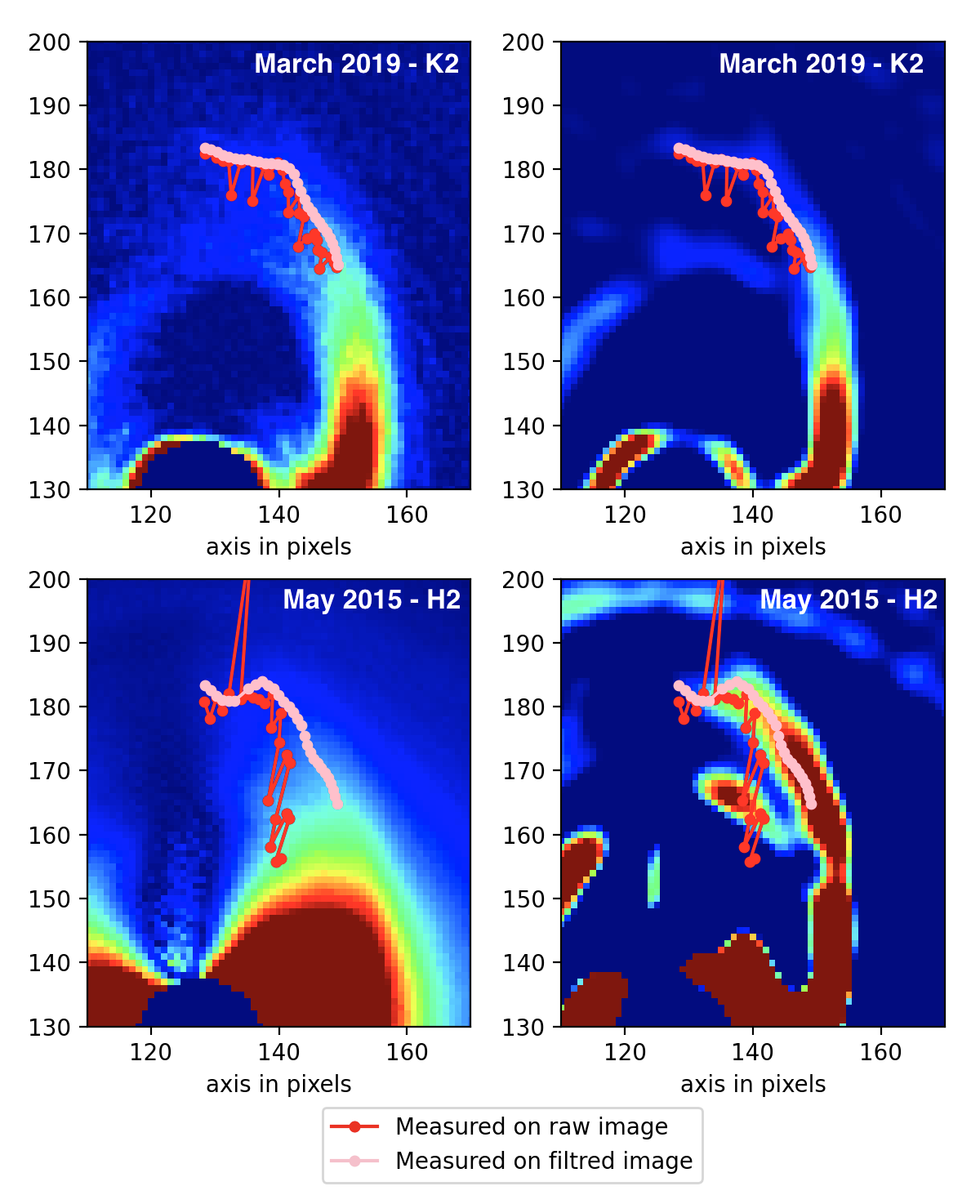}
    \caption{Demonstration of the positive effect of the Laplacian filter to help the identification of local radial maxima in two example data sets of different quality. The left column shows the raw image (after reduction, without filtering) and the right column shows the image after filtering. In both images the trace measurement is indicated.}
    \label{fig:laplace}
\end{figure}

\end{appendix}
\end{document}